\documentclass[article]{aa}
\usepackage{psfig,latexsym,longtable,lscape,rotating}
\begin{document}
\pagenumbering{arabic}
\title{Nearby early--type galaxies with ionized gas.I. \\ Line-strength indices of 
the underlying stellar population\footnote{Based on observations obtained at the  European Southern Observatory, La Silla, Chile (Programs Nr.~60.A-0647 and 61.A-0406}}
\author{Rampazzo R.$^1$, Annibali F.$^2$, Bressan A.$^{1,2}$, Longhetti M.$^3$, Padoan F.$^3$, Zeilinger W.W.$^4$, }
\institute{
 $^1$ INAF Osservatorio Astronomico di Padova, vicolo dell'Osservatorio 5, 35122 Padova, Italy\\
 $^2$ SISSA, via Beirut 4 - 34014 Trieste - Italy \\
$^3$ INAF Osservatorio Astronomico di Brera, via Brera 28, I-20121
 Milano, Italy \\
$^4$Institut f\" ur Astronomie der Universit\" at  Wien, T\" urkenschanzstra$\ss$e 17, A-1180 Wien, Austria 
}
 \offprints{R.~Rampazzo}
 \mail{rampazzo@pd.mi.astro.it}
\date{Received date; accepted date}
\authorrunning{Rampazzo et al.}
\titlerunning{Nearby early--type galaxies with ionized gas.I.}
\abstract{With the aim of building a data-set of spectral properties
of well studied  early-type galaxies showing emission lines, 
we present intermediate resolution spectra of 50 galaxies in the nearby Universe. 
The sample, which covers several of the E and S0 morphological
sub-classes, is biased toward objects that might be expected to have ongoing 
and recent star formation, at least in small amounts, because of the presence
of the emission lines. The emission are expected to come from the combination
of active galactic nuclei and star formation regions within the galaxies. 
Sample galaxies are located in environments corresponding to a broad range of 
local galaxy densities, although predominantly in low density environments.

Our long--slit spectra cover the 3700 - 7250 \AA\ wavelength range with a
spectral resolution of $\approx$7.6~\AA\ at 5550~\AA.  The specific aim of this
paper,  and our first step on the investigation, is to map the underlying galaxy
stellar population by measuring,  along the slit, positioned along the galaxy
major axis, line--strength indices at several, homogeneous galacto-centric
distances.  

For each object we extracted 7 luminosity weighted apertures (with radii: 
1.5\arcsec, 2.5\arcsec, 10\arcsec,  r$_e$/10, r$_e$/8, r$_e$/4 and r$_e$/2) 
corrected for the galaxy ellipticity and  4 gradients 
(0 $\leq$ r $\leq$r$_e$/16, r$_{e}$/16 $\leq$ r $\leq$r$_e$/8, r$_{e}$/8 $\leq$ 
r $\leq$r$_e$/4 and r$_{e}$/4 $\leq$ r $\leq$r$_e$/2). For each aperture and 
gradient we measured 25 line--strength indices: 21 of the set defined by the 
Lick-IDS ``standard'' system (Trager et al. \cite{Tra98}) and 4  introduced by 
Worthey \& Ottaviani  (\cite{OW97}). Line--strength indices have been transformed 
to the Lick-IDS system. Indices derived then include H$\beta$, Mg$_1$, Mg$_2$, 
Mgb, MgFe, Fe5270, Fe5335 commonly used in classic index-index diagrams. 

The paper introduces the sample, presents the observations, describes 
the data reduction procedures, the extraction of apertures and
gradients, the determination and correction of the line--strength
indices, the  procedure adopted to transform them into the Lick-IDS
System and the procedures adopted for the emission correction. We
finally discuss the comparisons between our dataset and line-strength
indices available in the literature. 

A significant fraction, about 60\%, of galaxies in the present sample
has one previous measurement in the Lick--IDS system but basically 
restricted within the r$_e$/8 region. Line-strength measures  obtained
both from apertures and gradients outside this area and within the
r$_e$/8 region, with the present radial mapping, are completely new. 

 }
\maketitle
\section{Introduction}
Ellipticals (Es) are among the most luminous and massive galaxies in the
Universe. Together with lenticular (S0) galaxies, composed of a bulge,
a stellar disk and often a stellar bar component they form the vast
category of early-type galaxies.
Although, Es appear as a uniform class of galaxies, populating a planar
distribution (the so-called Fundamental Plane) in the logarithmic
parameter space defined by the central stellar velocity dispersion
$\sigma$, the effective radius r$_e$ and effective surface brightness
I$_e$ (see e.g. Djorgovski \& Davis \cite{DD87}), much
evidence suggests that a secondary episode of star formation has occurred during 
their evolutionary history. Simulations indicate that galaxy collisions, 
accretion and merging episodes are important factors
in the evolution of galaxy shapes (see e.g. Barnes \cite{Ba96};
Schweizer \cite{Sch96}) and can interfere with their passive evolution.
This understanding of early-type galaxy formation has been
enhanced by the study of interstellar
matter. This component and its relevance in  secular galactic
evolution was widely neglected in early studies of early-type
galaxies since they were for a long time considered to be essentially
devoid of interstellar gas. In the last two decades, however, multi-wavelength
observations have changed this picture and have detected the presence of a
multi-phase Inter Stellar Medium (ISM): a hot (10$^7$ K) and a warm
(10$^4$ K) phase coexist and possibly interplay in several giant
ellipticals. A cool ($\approx$ 10 K) phase, detected in HI and
CO, is also often revealed in early-type galaxies (Bettoni et
al. 2001 and reference therein).
Unlike spiral galaxies, the bulk of the gas in ellipticals is heated
to the virial temperature, emitting in X-rays, and only comparatively
small quantities are detected in the warm and cool phase of the
interstellar medium (Bregman et al. \cite{Bre92}). 
The amount of X-ray emitting gas is related to the
optical luminosity of the galaxy (White \& Sarazin \cite{Wh91}), while
no relation is found between the properties of
the other components and the galaxy stellar
luminosity. This latter fact may indicate that some amounts of the
gaseous material may indeed be of external origin.
A pioneering spectroscopic study (Phillips et al. \cite{Phi86}), that
examined the properties of a set of 203 southern E and S0 galaxies,
began to shed light on the physical condition of the ionized gas in
early-type galaxies. On the grounds of their analysis of the [NII 6583]/H$\alpha$ 
ratio they pointed out that, when line emission is present, it is confined to
the nucleus and the properties of giant ellipticals are
``indistinguishable'' from a LINER nucleus (Heckman \cite{He80}).
When the ionized region is imaged using narrow band filters
centered at H$\alpha$+[NII 6583], it appears extended with
morphologies ranging from regular, disk-like structures to
filamentary structures (see e.g.  Ulrich-Demoulin \cite{U84}; Buson et
al. \cite{Bus93}; Zeilinger et al. \cite{Z96}; Macchetto et 
al. \cite{Mac96}) of several kpc in radius.  The gaseous disks appear to be
generally misaligned with respect to the stellar body of the galaxy suggesting
an external origin for most of the gaseous
matter. This picture is also supported by the observations of decoupled
kinematics of gas and stars in a significant fraction of early-type
galaxies (Bertola et al. \cite{BBZ92a}). Notwithstanding the large
amount of studies, our current understanding of the origin and the
nature of the ``warm'' ionized gas in elliptical galaxies is still rather
uncertain. The fact that the emission regions are always associated
with dust absorption, even in the  brightest X-ray systems, seems to
exclude ``cooling flows'' as the origin of the ionized gas (see
Goudfrooij \cite{Gou98}). The main ionization mechanism, which does not 
seem to be powered by star formation, however remains uncertain. 
Ionization mechanisms suggested range from photoionization by old
hot stars -- post--AGB and/or AGB-Manqu\'e type objects (Binette et al.
\cite{Bin94}) -- or mechanical energy flux from electron conduction in
hot, X-ray emitting gas (Voit \cite{Voi91}). Also ionization by a non-thermal 
central source is considered.
 
The present paper is the first of a series presenting a study of
early-type galaxies in the nearby Universe showing emission lines in
their optical spectra. Our aim is to improve the understanding
of the nature of the ionized gas in early-type galaxies by studying its
physical conditions, the possible ionization mechanisms, 
relations with the other gas components of the ISM and the connection
to the stellar population of the host galaxy. The adopted
strategy is to investigate galaxy spectra of intermediate spectral
resolution at different galactocentric distances and to attempt the modeling
of their stellar populations to measure emission line properties. The study of
stellar populations of early-type galaxies is of fundamental importance to the
understanding of their evolution by the measurement of the evolution of the 
spectral energy distribution with time (see e.g.  Buzzoni et al. \cite{Buz92};
Worthey \cite{Wor92};  Gonz\' alez \cite{G93}; Buzzoni et al. \cite{Buz94}; 
Worthey et al.\cite{Wor94}; Leonardi \& Rose \cite {LR96}; Wothey \& Ottaviani
\cite{OW97}; Trager et al.~\cite{Tra98}; Longhetti et al. \cite{L98a}; Vazdekis
\cite{Vaz99}; Longhetti et al. \cite{L99}; Longhetti et al. \cite{L00}; Trager et
al. \cite{Tra00}; Kuntschner et al. \cite{Ku00}; Beuing et al.~\cite{Beu02}; 
Kuntschner et al. \cite{Ku02}; Thomas et al. ~\cite{Thom03}; Mehlert et al. 
\cite{Mehl03}) . Investigating issues such as the evolution of stellar
populations and the ISM, we will explore the complex, evolving ecosystem within
early--type galaxies and build a database of well studied galaxies to be used as
a reference set for the study of intermediate and distant objects.  Our
target is to characterize the stellar populations, in particular those
related to the extended emission region, in order to constrain
hints about the galaxy formation/evolution history from the modeling
of the complete (lines and continuum) spectrum characteristics.
In this paper we present the sample, the observations and the data
reduction and we discuss, through the comparison with the literature,
the database of line--strength indices we have measured. Forthcoming
papers will analyze the emission region by  tracing the properties as a
function of the distance from the galaxy center.
The paper is organized as follows. Section~2 introduces the sample and some
of the relevant properties useful to infer the ionized gas origin and
nature. Section~3 presents the observations, the data reduction and
the criteria and the methods used for the selection of apertures and 
gradients extracted from the long slit spectra. Section~4 details the 
transformation of the line-strength indices to the Lick-IDS System
and provides the database of line--strength indices measured at different galactocentric
distances. In Section~5 we review the results, providing the database 
of line--strength indices measured at different galactocentric
distances and discuss the comparison with the literature. 
The Appendix A provides a description/comments of
individual galaxies in the sample.

\begin{table*}
\tiny{
\begin{tabular}{lcrllrrcrrclc}
& & & & & & & & & &  & &\\
\multicolumn{13}{c}{\bf Table 1 Overview of the observed sample} \\
\hline\hline
\multicolumn{1}{c}{ident}
& \multicolumn{2}{c}{R.A. (2000) Dec.}
& \multicolumn{1}{c}{RSA}
& \multicolumn{1}{c}{RC3}
& \multicolumn{1}{c}{P.A.}
& \multicolumn{1}{c}{B$_0$}
& \multicolumn{1}{c}{(B-V)$_0$}
 & \multicolumn{1}{c}{(U-B)$_0$}
& \multicolumn{1}{c}{V$_{hel}$}
& \multicolumn{1}{c}{r$_e$}
& \multicolumn{1}{c}{$\rho_{xyz}$} 
& \multicolumn{1}{c}{$\epsilon$} \\
 \hline
 NGC~128  & 00 29 15.1 &  02 51 50 & S02(8) pec   & S0 pec sp &   1    & 12.63 & 0.87 & 0.51 & 4227 &  17.3 &   & 0.67    \\
 NGC~777  & 02 00 14.9 &  31 25 46 & E1                & E1             & 155  & 12.23 & 0.99 &         & 5040 &  34.4 & & 0.21 \\
 NGC~1052 & 02 41 04.8& -08 15 21 & E3/S0           & E4            & 120  & 11.53 & 0.89 & 0.44 & 1475 & 33.7  & 0.49 & 0.28  \\
 NGC~1209 & 03 06 03.0 & -15 36 40 & E6               & E6:           &  80   & 12.26 & 0.90 & 0.40 & 2619  & 18.5  & 0.13 & 0.52 \\
 NGC~1297 & 03 19 14.2 & -19 05 59 & S02/3(0)      & SAB0 pec: &   3  & 12.61 &         &         & 1550 &   28.4 &  0.71  & 0.13   \\
 NGC~1366 & 03 33 53.7& -31 11 39 & E7/S01(7)     & S0 sp        &   2  & 12.81 &         &         & 1310 &   10.6  & 0.16 & 0.56     \\
 NGC~1380 & 03 36 27.3 & -34 58 34 & S03(7)/Sa    & SA0           &   7  & 11.10 & 0.90 & 0.44 & 1844 &   20.3 & 1.54   & 0.41   \\
 NGC~1389 & 03 37 11.7& -35 44 45 & S01(5)/SB01 & SAB(s)0-:  &  30 & 12.39 & 0.89 & 0.39 &  986 &    15.0 & 1.50  & 0.37     \\
 NGC~1407 & 03 40 11.8 & -18 34 48 & E0/S01(0)    & E0            &  35 & 10.93 & 0.89 &      & 1766    & 70.3 & 0.42 & 0.07 \\
 NGC~1426 & 03 42 49.1 & -22 06 29 & E4                & E4            & 111 & 12.37 & 0.80 & 0.34 & 1443 &  25.0    & 0.66 & 0.34\\
 & & & & & & & & & &  & & \\
 NGC~1453 & 03 46 27.2 & -03 58 08 & E0                 & E2            &  45 & 12.59 & 0.92 & 0.56 & 3906 &    25.0   &  & 0.17\\
 NGC~1521 & 04 08 18.9 & -21 03 07 & E3                 & E3            &  10 & 12.58 & 0.86 & 0.42 & 4165 &    25.5    & & 0.35 \\
 NGC~1533 & 04 09 51.9& -56 07 07 & SB02(2)/SBa  & SB0-        & 151 & 11.65 & 0.89 & 0.46 &  773 &    30.0  &  0.89& 0.19     \\
 NGC~1553 & 04 16 10.3& -55 46 51 & S01/2(5)pec   & SA(r)0      & 150 & 10.36 & 0.85 & 0.41 & 1280 &   65.6 & 0.97 & 0.38 \\
 NGC~1947 & 05 26 47.5& -63 45 38 & S03(0) pec     & S0- pec    & 119 & 11.75 & 0.93 & 0.44 & 1100 &   32.1 & 0.24  &0.11   \\
 NGC~2749 & 09 05 21.4 &  18 18 49 & E3                 & E3            &  60 & 13.03 & 0.87 & 0.44 & 4180 &     33.7    &  & 0.07\\
 NGC~2911 & 09 33 46.1 &  10 09 08 & S0p or S03(2)& SA(s)0: pec & 140 & 12.53 & 0.91 & 0.43 & 3131 & 50.9  &     & 0.32  \\
 NGC~2962 & 09 340 53.9 &  05 09 57 & RSB02/Sa   & RSAB(rs)0+ &   3 & 12.71 & 1.00 &      & 2117 &      23.3  &  0.15  & 0.37   \\
 NGC~2974 & 09 42 33.2& -03 41 55 & E4                 & E4            &  42 & 11.68 & 0.89 & 0.51 & 1890 &     24.4  & 0.26 & 0.38\\
 NGC~3136 & 10 05 47.9 & -67 22 41 & E4                 & E:            &  40 & 11.42 & 0.60 & 0.23 & 1731 &  36.9    & 0.11 & 0.24\\
 & & & & & & & & & &  & & \\
 NGC~3258 & 10 28 54.1& -35 36 22 & E1                 & E1            &  76 & 12.48 & 0.94 & 0.37 & 2778 &  30.0 & 0.72 & 0.13\\
 NGC~3268 & 10 30 00.6 & -35 19 32 & E2                & E2            &  71 & 12.57 & 0.94 & 0.41 & 2818 & 36.1       & 0.69 & 0.24\\
 NGC~3489 & 11 00 18.3 &  13 54 05 & S03/Sa          & SAB(rs)+      &  70 & 11.13 & 0.74 & 0.34 &  693 &  20.3  & 0.39  & 0.37   \\
 NGC~3557 & 11 09 57.5 & -37 32 22 & E3                 & E3            &  30 & 11.23 & 0.86 & 0.48 & 3038 & 30.0      & 0.28 & 0.21\\
 NGC~3607 & 11 16 54.3 &  18 03 10 & S03(3)          & SA(s)0:       & 120 & 11.08 & 0.88 & 0.43 &  934 &  43.4  & 0.34 & 0.11 \\
 NGC~3962 & 11 54 40.1& -13 58 31 & E1                 & E1            &  15 & 11.61 & 0.89 &      & 1822 & 35.2   & 0.32 & 0.22 \\
 NGC~4552 & 12 35 39.8&  12 33 23 & S01(0)           & E             &  92 & 10.80 & 0.95 & 0.54 &  322 & 29.3  & 2.97  & 0.09\\
 NGC~4636 & 12 42 50.0 &  02  41 17 & E0/S01(6)    & E0-1          & 150 & 10.50 & 0.89 & 0.46 &  937 &  88.5 & 1.33 & 0.24\\
 NGC~5077 & 13 19 31.6 & -12 39 26 & S01/2(4)       & E3+           &  11 & 12.52 & 0.98 & 0.54 & 2764 & 22.8  & 0.23 & 0.15\\
 NGC~5328 & 13 52 53.6& -28 29 16 & E4                 & E1:           &  87 & 12.78 & 0.73 &      & 4671 & 22.2              & & 0.31 \\
 & & & & & & & & & &  & &\\
 NGC~5363 & 13 56 07.1&  05 15 20 & [S03(5)]         & I0:           & 135 & 11.06 & 0.90 & 0.50 & 1138 &  36.1 & 0.28 & 0.34     \\
 NGC~5846 & 15 06 29.2 &  01 36 21 & S01(0)          & E0+           &   1 & 11.13 & 0.93 & 0.55 & 1709 &  62.7  & 0.84 & 0.07 \\
 NGC~5898 & 15 18 13.6 & -24 05 51 & S02/3(0)       & E0            &  30 & 12.41 & 0.95 &      & 2267 &  22.2 & 0.23 & 0.07\\
 NGC~6721  & 19 00 50.4 & -57 45 28 & E1               & E+:           & 155 & 12.93 & 0.87 & 0.45 & 4416 &    21.7  &  & 0.15\\
 NGC~6868 & 20 09 54.1 & -48 22 47 & E3/S02/3(3)  & E2            &  86 & 11.72 & 0.91 & 0.52 & 2854 & 33.7    & 0.47 &0.19 \\
 NGC~6875 & 20 13 12.4 & -46 09 38 & S0/a(merger) & SAB(s)0- pec: &  22 & 12.66 & 0.82 & 0.29 & 3121 &   11.7      &  & 0.41     \\
 NGC~6876 & 20 18 19.0 & -70 51 30 & E3                 & E3            &  80 & 12.45 & 0.91 & 0.57 & 3836 &  43.0  & & 0.13 \\
 NGC~6958 & 20 48 42.4 & -37 59 50 & R?S01(3)      & E+            & 107 & 12.13 & 0.82 & 0.40 & 2652 &   19.8  & 0.12 & 0.15\\
 NGC~7007 & 21 05 28.0 & -52 33 04 & S02/3/a        & SA0-:         &   2 & 12.92 & 0.91     & 0.40     & 2954 &   15.4   &  0.14   & 0.42  \\
NGC~7079 & 21 32 35.1 & -44 04 00 & SBa                & SB(s)0        &  82 & 12.49 & 0.77 & 0.25 & 2670 &        19.8    &  0.19 & 0.32\\
& & & & & & & & & &  & & \\
NGC~7097 & 21 40 13.0 & -42 32 14 & E4                 & E5            &  20 & 12.48 & 0.88 & 0.42 & 2404        &       18.1      & 0.26  & 0.29    \\
NGC~7135 & 21 49 45.5 & -34 52 33 & S01 pec        & SA0- pec   &  47 & 12.61 & 0.91 & 0.45 & 2718 &    31.4         &  0.32   & 0.31  \\
NGC~7192 & 22 06 50.3 & -64 18 56 & S02(0)          & E+:           &  -  & 12.15 & 0.88 & 0.48 & 2904 &    28.6       & 0.28 & 0.15\\
NGC~7332 & 22 37 24.5 &  23 47 54 & S02/3(8)        & S0 pec sp     & 155 & 11.58 & 0.76 & 0.25 & 1207 & 14.7  & 0.12  & 0.75  \\
NGC~7377 & 22 47 47.4 & -22 18 38 & S02/3/Sa pec & SA(s)0+       & 101 & 12.61 & 0.91 & 0.29 & 3291 &  36.9  &     & 0.19  \\
IC~1459 & 22 57 10.6 & -36 27 44& E4                    & E             & 40 & 10.96 & 0.90 & 0.54 & 1659 &   34.4  & 0.28  & 0.28\\
IC~2006 & 03 54 28.5& -35 57 58 & E1                    & E             &  -  & 12.27 & 0.90 & 0.42 & 1350 &  28.6           & 0.12 & 0.15\\
IC~3370 & 12 27 38.0 & -39 20 17 & E2 pec             & E2+           &  45 & 11.91 & 0.91 & 0.35 & 2934 &    38.6     & 0.20 & 0.21\\
IC~4296 & 13 36 39.4 & -33 58 00 & E0                    & E             &  40 & 11.43 & 0.90 & 0.54 & 3762 & 41.4   &  & 0.17\\
IC~5063 & 20 52 02.4 & -57 04 09 & S03(3)pec/Sa   & SA(s)0+:      & 116 & 13.14 & 0.91 & 0.26 & 3402 &      26.7   &   & 0.28    \\
\hline
\end{tabular}}
\label{table1}

\medskip
{{\bf Notes}: The value of r$_e$ of NGC 1297, NGC 6876 have been derived from ESO-LV (Lauberts \& Valentijn (\cite{LV89}).}
\end{table*}
\begin{figure*}
\resizebox{8.5cm}{!}
{\psfig{figure=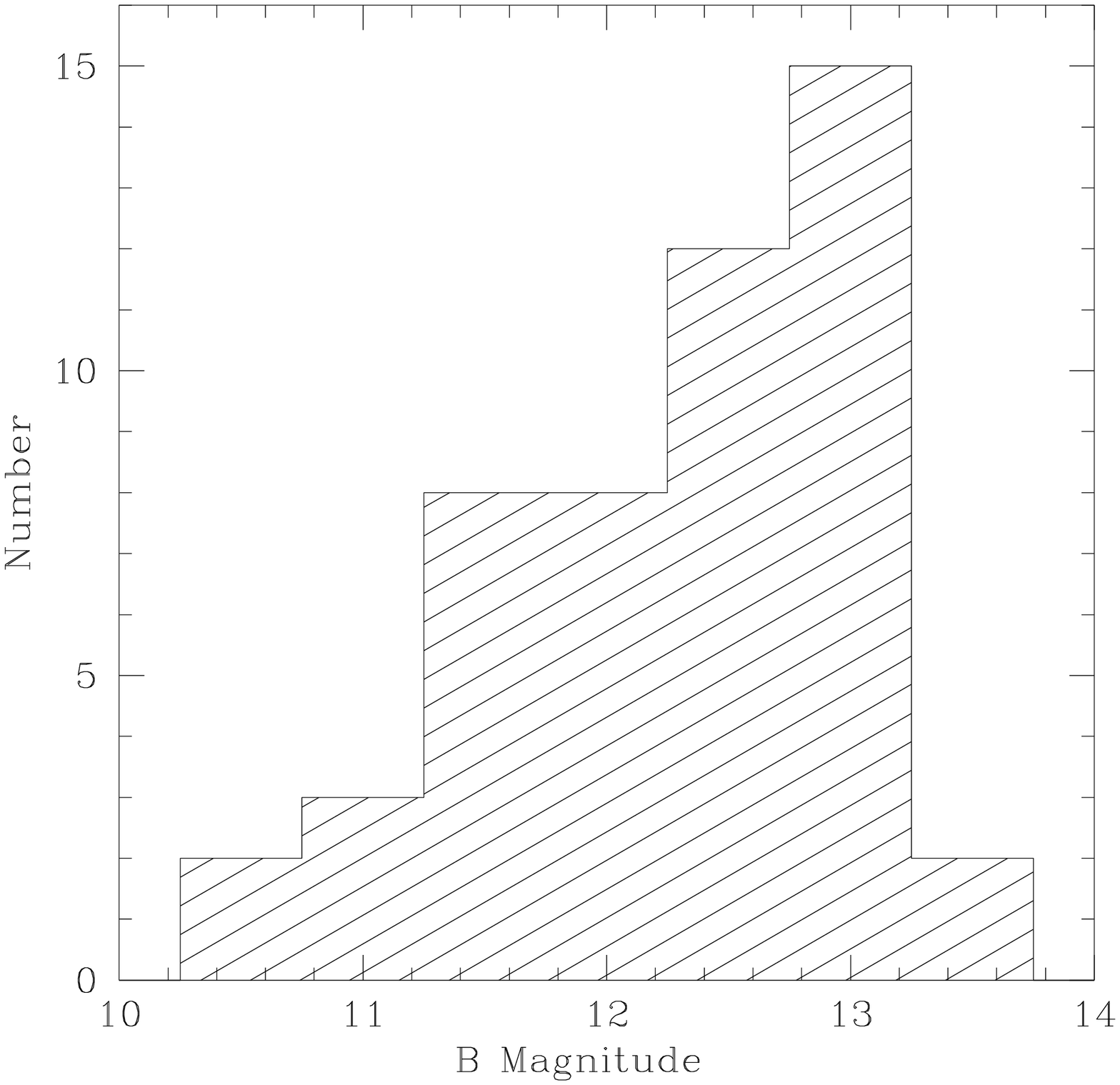,width=16cm,clip=}
\psfig{figure=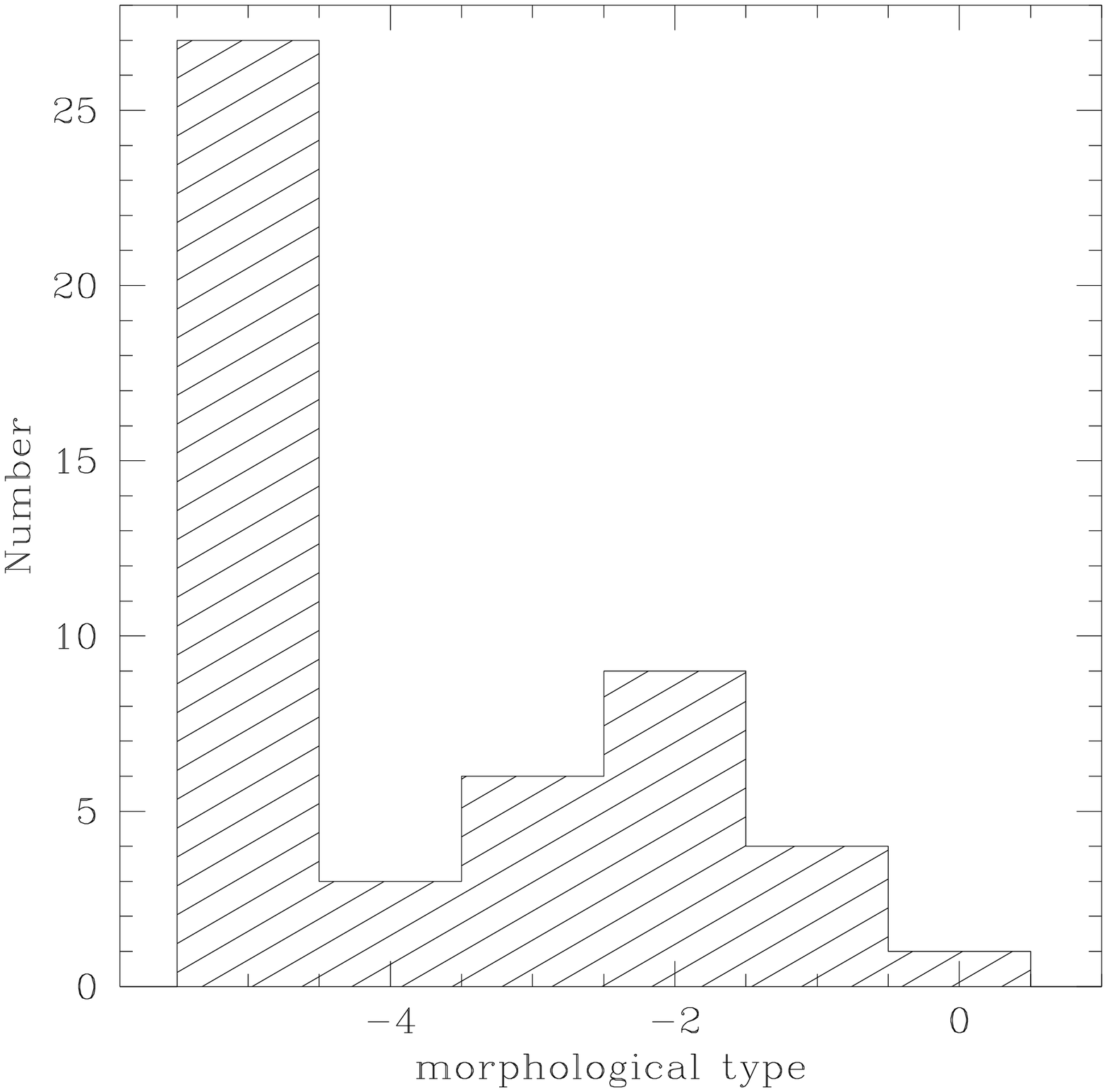,width=16cm,clip=} }
\resizebox{8.5cm}{!}
{ \psfig{figure=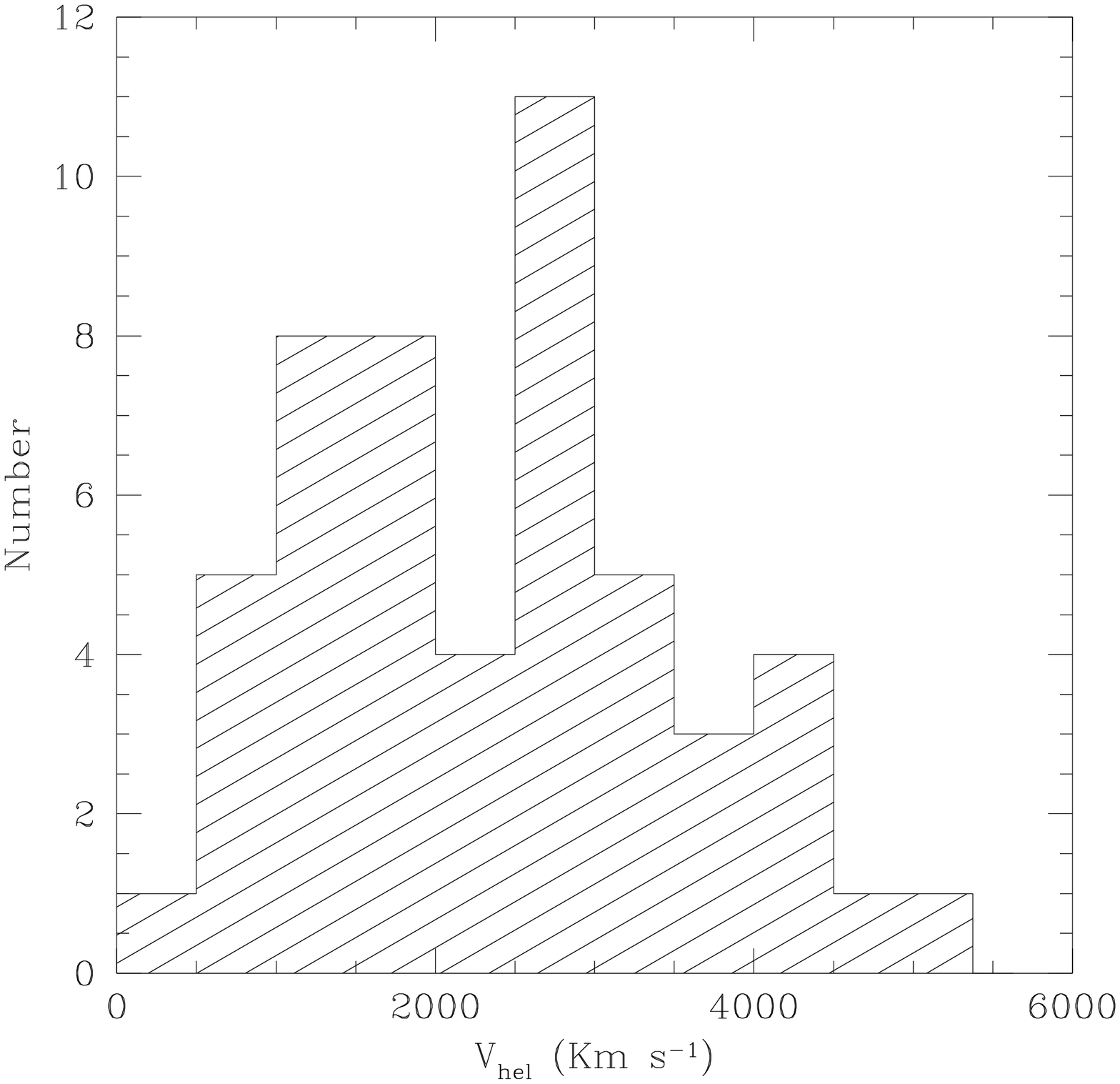,width=16cm,clip=}
\psfig{figure=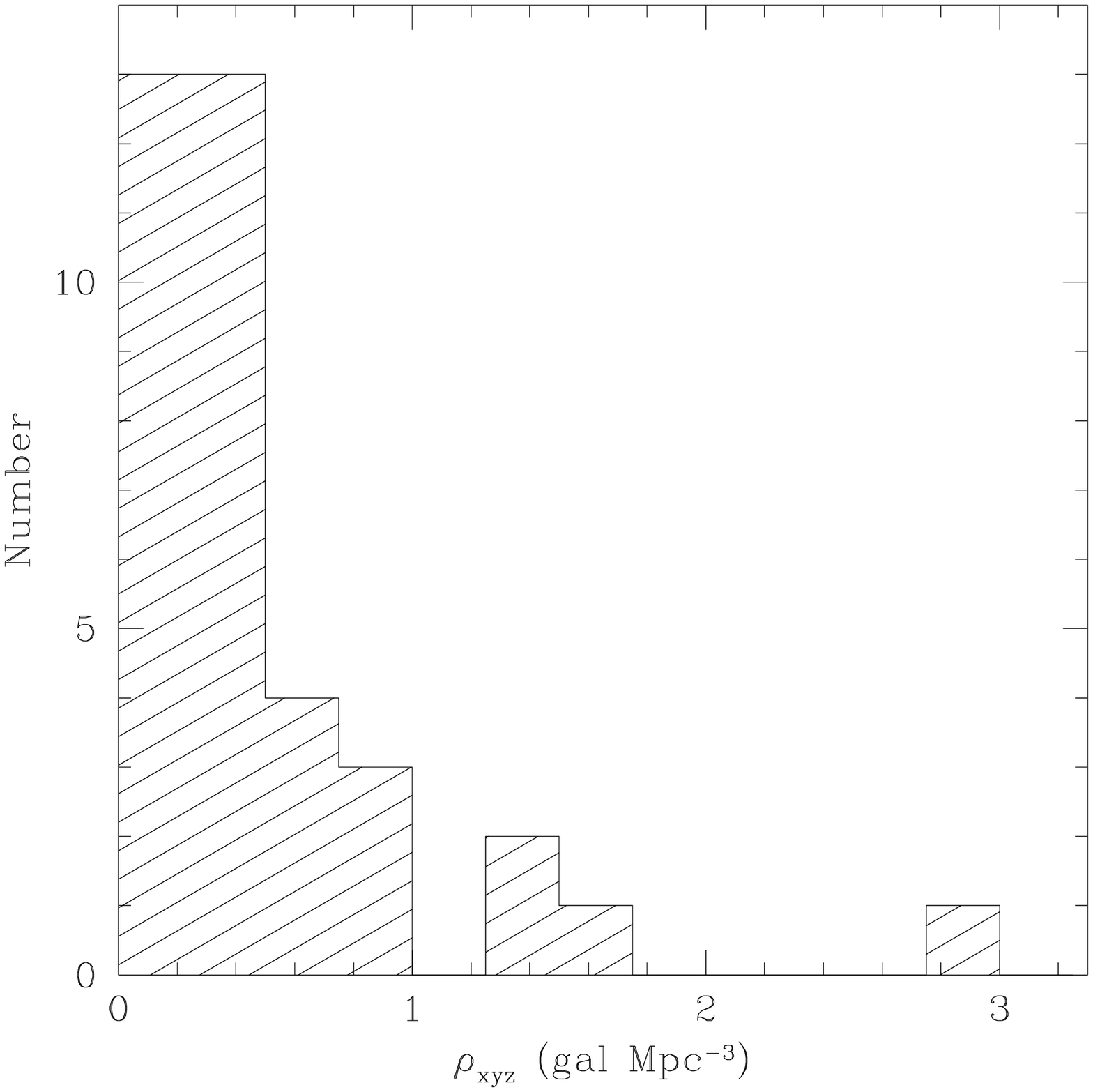,width=16cm,clip=} } 
\caption{Distribution of B-magnitudes (first panel), morphological types (second panel), 
heliocentric velocity (third panel) and galaxy density (forth panel).} 
\label{fig1}
\end{figure*}

\begin{figure*}
\resizebox{16cm}{!}
{ \psfig{figure=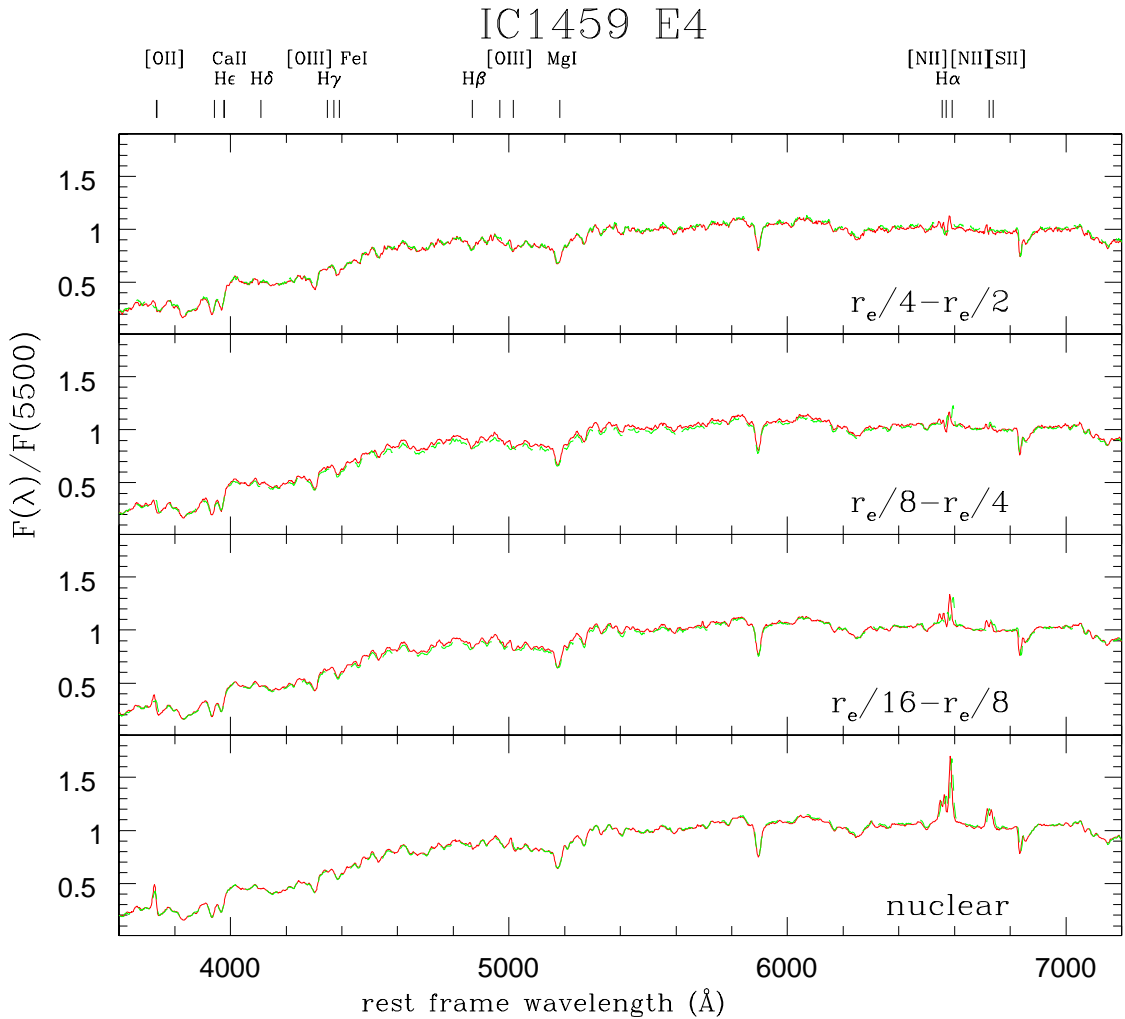,width=16cm,clip=}}
\caption{Spectra of a representative galaxy in the sample. The figure shows the gradients
obtained sampling each long--slit spectrum in four regions: between 
0 $\leq$ r $\leq$r$_e$/16 (indicated as "nuclear" in the figure), r$_{e}$/16 $\leq$ r $\leq$r$_e$/8,
r$_{e}$/8 $\leq$ r $\leq$r$_e$/4 and r$_{e}$/4 $\leq$ r $\leq$r$_e$/2. 
For each region the figure overplots the two opposite sides of the galaxies with
respect to the nucleus. With the exclusion of few cases (see text), the two 
sides agree within few percent. The major difference between
the two sides often reside in the emission line distribution which is not symmetric 
with respect to the nucleus.}
\label{fig2}
\end{figure*}

\section{Characterization of the sample}
Our sample contains  50 early--type galaxies. The sample is selected
from a compilation of galaxies showing ISM traces in at least one of the
following bands: IRAS 100 $\mu$m, X-ray, radio, HI and CO (Roberts et
al. \cite{Ro91}).  All galaxies belong to {\it Revised Shapley Ames
Catalog of Bright Galaxies (RSA)} (Sandage \& Tammann \cite{RSA}) and
have a redshift of less than 5500 km~s$^{-1}$. The sample should then be
biased towards objects that might be expected to have ongoing and recent
star formation, at least in small amounts, because of the presence of emission
lines. The emission should come from a combination of active galactic 
nuclei and star formation regions within the galaxies.
Table~\ref{table1} summarizes  the basic characteristics of the
galaxies available from the literature.  Column (1)
provides the identification; column (2) and (3) the R.A. \& Dec.
coordinates; column (4) and (5) the galaxy morphological classifications 
according to the RSA (Sandage \& Tamman \cite{RSA}) and RC3 (de Vaucouleurs et
al. \cite{RC3}) respectively. Columns (6), (7), (8) (9) give the position angle of
the isophotes along major axis, the total corrected magnitude and the total
(B-V) and (U-B) corrected colors from RC3 respectively. The
heliocentric systemic velocity from HYPERCAT
({\tt http://www-obs.univ-lyon1.fr/hypercat}) is reported in column
(10). The effective radius, derived from A$_e$,  the diameter of the
effective aperture from RC3, is given in column (11). A measure of
the richness of the environment, $\rho_{xyz}$ (galaxies~Mpc$^{-3}$), 
surrounding each galaxy  is reported in column (12) (Tully
\cite{Tu88}). Column (13) lists the average ellipticity of the
galaxy as obtained from HYPERCAT.  

Figure~\ref{fig1} summarizes the basic characteristics
of the present sample,  and, in particular  in the fourth panel,
provides evidence that a large fraction of galaxies are in low density environments.
In the following subsection we summarize morphological
and photometric studies of the ionized component which provide an insight of
the overall galaxy structure.  In the Appendix A we complement the above
information with individual notes on galaxies emphasizing kinematical studies of
the ionized gas component, its correlation with the stellar body and
its possible origin.

\subsection{Imaging surveys of the ionized gas component}
Buson et al. (\cite{Bus93}) presented  H$\alpha$+[NII] imaging of a set of 15
nearby elliptical and S0 galaxies with extended optical emission regions.
Nine are included in our sample, namely NGC~1453, NGC~1947, NGC~2974,
NGC~3962, NGC~4636, NGC~5846, NGC~6868, NGC~7097 and IC~1459. In most
of these galaxies the extended emission forms an inclined disk with ordered
motions (Zeilinger et al. \cite{Z96}).
Furthermore the major axes of the stellar and the gaseous components
appear frequently misaligned. In NGC~1453 the emission appears strongly
decoupled from the stellar component and roughly aligned with the
minor axis of the galaxy. In NGC~1947 the emission appears associated
with a complex systems of dust lanes and shows several distinct knots. In NGC~2974
the ionized gas appears to lie in a fundamentally regular elongated
structure with some peripheral fainter filaments (dust is also
present).  Buson et al. (\cite{Bus93}) reported that for NGC~3962 the
emission region consists of two distinct subsystems: an
elongated central component strongly misaligned with both the major
and minor axes of the stellar figure, and a peculiar extended arm-like
structure departing from the major axis of the internal disk,
crossing the stellar body at an angle of 180$^\circ$. Some dust is
also noted. NGC~4636 has a ring-like emitting region extending
asymmetrically around the galaxy nucleus. NGC~5846 shows complex,
filamentary emission-line morphology, including an arm-like feature
and dusty patches. NGC~6868 shows an elongated emitting region, with
faint peripheral extensions and dust patches, which are strongly
decoupled from the stellar component. NGC~7079 has elongated
emission misaligned by about 30$^\circ$ with the stellar figure
associated with dusty features. The ionized region of IC~1459 is
aligned with the stellar major axis. Dust features for this latter
galaxy have been identified by Goudfrooij (\cite{Gou94}).
A characterization of the extended emission region was attempted
by Macchetto et al. (\cite{Mac96}) who observed 73 luminous
early--type galaxies selected from the RC3 catalogue.  CCD images were
obtained in broad R-band and narrow band images centered on the
H$\alpha$ + [NII] emission lines. A set of 17 galaxies of our sample
galaxies are in the Macchetto et al.  set, namely NGC~1407, NGC~1453,
NGC~3489, NGC~3607, NGC~3268, NGC~4552, NGC~4636, NGC~5077, NGC~5846,
NGC~5898, NGC~6721, NGC~6868, NGC~6875, NGC~6876, NGC~7192, IC~1459
and IC~4296.
The morphology of the extended emission has been divided into three
categories. ``SD'' indicates galaxies with a small disk, extended on average  
less than 4 kpc (they adopt H$_0$=55 km s$^{-1}$ Mpc$^{-1}$) with
sometimes faint and short filaments. The ionized regions of NGC~1407, NGC~3268, 
NGC~3489, NGC~3607, NGC~4552 and NGC~4636  were classified as ``SD''.
``RE'' indicates regular extended regions, similar to  previous ones but  
extended between 4 and 18 kpc. NGC~1453, NGC~5077, NGC~5846, NGC~5898, 
NGC~6721, NGC~6868, NGC~6875, NGC~7192 and IC~1459 have extended ionized regions.
``F'' represents the detection of filaments and conspicuous filamentary
structure which dominate the morphology and which depart from a more regular
disk--like inner region. The filamentary structures extend as far as 10 kpc from
the galaxy center. However, our sample contains no galaxies with a filamentary
morphology. Macchetto and co-authors did not classify the ionized regions of
NGC~6876 and IC~4296. From the above notes it appears that the ionized region of
a large fraction of galaxies  in our sample have  a disk--like structure. A
large number of kinematical studies (see individual notes in Appendix A) supports
this hypothesis. In addition, it emerges that the ionized
emission is always associated with dust absorption, even in the brightest X-ray
systems (Goudfrooij \cite{Gou94}; Goudfrooij \cite{Gou98}). Several
classes of galaxies are present in the sample: interacting or post-interacting
galaxies, galaxies showing evidence of kinematical decoupling between galaxy
sub-components, elliptical galaxies with a dust lane along the minor axis, radio
galaxies and galaxies hosting an AGN nucleus. 
To summarize the individual notes in Appendix A: (a) the sample contains four
galaxies showing a shell structure (namely NGC~1553, NGC~4552, NGC~6958,
NGC~7135).  Twenty galaxies (namely NGC~128, NGC~1052, NGC~1407, NGC~1947,
NGC~2749, NGC~3136, NGC~3489, NGC~4636, NGC~5077, NGC~5363, NGC~5846, NGC~5898,
NGC~6868, NGC~7007, NGC~7097, NGC~7192, NGC~7332, IC~1459, IC~2006, IC~4296)
have a peculiar kinematical behavior, i.e. rotation along the apparent minor
axis, turbulent gas motions, counterrotation of star vs. gas and/or star vs. stars. In four galaxies (namely
NGC~1052, NGC~1553, NGC~3962 and NGC~7332) multiple gas components have been
detected.
\begin{table*}
\small{
\begin{tabular}{lll}
& &   \\
\multicolumn{3}{c}{\bf Table 2. ~~~Observing parameters} \\
\hline\hline
\multicolumn{1}{c}{}
& \multicolumn{1}{l}{Run 1}
& \multicolumn{1}{l}{Run 2}
 \\
\hline
Date of Observations & March 98 & September 98 \\
Observer  & Zeilinger W.& Zeilinger W. \\
Spectrograph & B \& C grating \#25& B \& C grating \#25 \\
Detector & Loral 2K UV flooded & Loral 2K UV flooded \\
Pixel size ($\mu$m) & 15  & 15 \\
Scale along the slit (\arcsec/px$^{-1}$) & 0.82  & 0.82  \\
Slit length (\arcmin) & 4.5 & 4.5 \\
Slit width (\arcsec) & 2 & 2 \\
Dispersion(\AA\ mm$^{-1}$) & 187 & 187 \\
Spectral Resolution (FWHM at 5200 \AA\ ) (\AA ) & 7.6 & 7.6  \\
Spectral Range (\AA ) & 3550-9100 & 3550-9100 \\
Seeing Range(FWHM) (\arcsec) & 1.2-2 & 1.0-2.0 \\
Standard stars & Feige 56 & ltt 1788, ltt 377 \\
\hline
\end{tabular}}
\label{table2}
\end{table*}
\begin{table*}
\small{
\begin{tabular}{lccccccc}
& & & & & & &  \\
\multicolumn{8}{c}{\bf Table 3. ~~~Journal of galaxy observations} \\
\hline\hline
\multicolumn{1}{c}{ident.}&
\multicolumn{1}{c}{Run} &
\multicolumn{1}{c}{Slit PA} &
\multicolumn{1}{c}{t$_{exp}$} &
\multicolumn{1}{c}{ident.} &
\multicolumn{1}{c}{Run} &
\multicolumn{1}{c}{Slit PA} &
\multicolumn{1}{c}{t$_{exp}$ }   \\
\multicolumn{1}{c}{}& 
\multicolumn{1}{c}{} &
\multicolumn{1}{c}{[deg]} & 
\multicolumn{1}{c}{[sec]} &
\multicolumn{1}{c}{} & 
\multicolumn{1}{c}{} & 
\multicolumn{1}{c}{[deg]} & 
\multicolumn{1}{c}{[sec]}  \\
\hline
NGC 128  & 2&   1  & 2$\times$1800& NGC 3962 & 1 &  15& 2$\times$1800\\
NGC 777  & 2& 155& 2$\times$1800& NGC 4552 & 1 &  92& 2$\times$1800 \\
NGC 1052 & 2& 120& 2$\times$1800& NGC 4636 & 1 & 150& 2$\times$1800 \\
NGC 1209 & 2&  80& 1$\times$2400& NGC 5077 & 1 &  11& 2$\times$1800 \\
NGC 1297 & 2&   3& 2$\times$1800& NGC 5328 & 1 &  87& 2$\times$1800 \\
NGC 1366 & 2&   2& 2$\times$1800& NGC 5363 & 1 & 135& 2$\times$1800 \\
NGC 1380 & 2&   7& 2$\times$1800& NGC 5846 & 1 &   1& 2$\times$1800 \\
NGC 1389 & 2&  30& 2$\times$1800& NGC 5898 & 1 &  30& 2$\times$1800 \\
NGC 1407 & 2&  35& 2$\times$1800& NGC 6721 & 1,2&155& 4$\times$1800 \\
NGC 1426 & 2& 111& 2$\times$1800& NGC 6868 & 2 &  86& 2$\times$1800 \\
 & & & &  & & &   \\
NGC 1453 & 2&  45& 2$\times$1800& NGC 6875 & 2&  50&3$\times$1800  \\
NGC 1521 & 2&  10& 2$\times$1800& NGC 6876 & 2&  75&2$\times$1800  \\
NGC 1533 & 2& 151& 1$\times$2400& NGC 6958 & 2& 107&2$\times$1800  \\
NGC 1553 & 2& 150& 2$\times$1800& NGC 7007 & 2&   2&2$\times$1800  \\
NGC 1947 & 2&  29& 2$\times$1800& NGC 7079 & 2&  82&2$\times$1800\\
NGC 2749 & 1&  60& 2$\times$1800& NGC 7097 & 2&  20&2$\times$1800  \\
NGC 2911 & 1& 140& 2$\times$1800& NGC 7135 & 2&  47&2$\times$1800  \\
NGC 2962 & 1&   3& 1$\times$1800& NGC 7192 & 2&  90&2$\times$1800  \\
NGC 2974 & 1&  42& 2$\times$1800& NGC 7332 & 2& 155&2$\times$1800 \\
NGC 3136 & 1&  40& 2$\times$1800& NGC 7377 & 2& 101&2$\times$1800  \\
 & & & &  &  & &  \\
NGC 3258 & 1&  76& 2$\times$1800& IC 1459 & 2&   40&2$\times$1800  \\
NGC 3268 & 1&  71& 2$\times$1800& IC 2006 &  2&  45&2$\times$1800 \\
NGC 3489 & 1&  70& 2$\times$1800& IC 3370 & 1&   45&2$\times$1800 \\
NGC 3557 & 1&  30& 2$\times$1800& IC 4296 & 1&   40&2$\times$1800  \\
NGC 3607 &1 &120& 2$\times$1800& IC 5063 & 2&  116&2$\times$1800   \\
\hline
\end{tabular}}
\label{table3}
\end{table*}
\begin{table*}
\small{
\begin{tabular}{lcccllcccl}
& & & & & & &  \\
\multicolumn{8}{c}{\bf Table 4. ~~~ Velocity dispersion values adopted in the correction of line--strength indices} \\
\hline\hline
\multicolumn{1}{l}{Ident.} &
\multicolumn{1}{c}{$\sigma_{r_e/8}$} &
\multicolumn{1}{c}{$\sigma_{r_e/4}$} &
\multicolumn{1}{c}{$\sigma_{r_e/2}$} &
\multicolumn{1}{l}{Ref.} &
\multicolumn{1}{l}{Ident.} &
\multicolumn{1}{c}{$\sigma_{r_e/8}$} &
\multicolumn{1}{c}{$\sigma_{r_e/4}$} &
\multicolumn{1}{c}{$\sigma_{r_e/2}$} &
\multicolumn{1}{l}{Ref.} \\
\multicolumn{1}{c}{} &
\multicolumn{1}{c}{[km~s$^{-1}$]} &
\multicolumn{1}{c}{[km~s$^{-1}$]} &
\multicolumn{1}{c}{[km~s$^{-1}$]} &
\multicolumn{1}{c}{} &
\multicolumn{1}{c}{} &
\multicolumn{1}{c}{[km~s$^{-1}$]} &
\multicolumn{1}{c}{[km~s$^{-1}$]} &
\multicolumn{1}{c}{[km~s$^{-1}$]} &
\multicolumn{1}{c}{}\\
\hline
NGC 128   &183& --  & --  &      & NGC 3962  &225&  --   & --    &       \\
NGC 777   &317& 272 & 266 & JS89 & NGC 4552  &264& 226 & 214 & SP97b  \\
NGC 1052  &215& 179 & 179 & FI94 & NGC 4636  &209& 202 &     & CMP00  \\
NGC 1209  &240& 195 & 178 & PS98 & NGC 5077  &260& 239 & 228 & CMP00  \\
NGC 1297  & 115& --  &  -- &      & NGC 5328  &303&   -- &  -- &       \\
NGC 1366  & 120& --  &  -- &      & NGC 5363  &199& 181  & 148 & S83   \\
NGC 1380  &240& 220 & 198 & DO95 & NGC 5846  &250& 228  & 190 & CMP00 \\
NGC 1389  & 139&  -- & --  &      & NGC 5898  &220& 183  & 172 & CMP00 \\
NGC 1407  &286&  -- & --  &      & NGC 6721  &262& 245  & 171 & B94 \\
NGC 1426  &162*& 157 & 121& PS97a& NGC 6868  &277& 235  & 220 & CMP00\\
 & & & & & & & & &    \\
NGC 1453  &289&  -- &  -- &      & NGC 6875  & --&  --  &  -- &    \\
NGC 1521  &235& 236 & 206 & PS98 & NGC 6876  &230&  --  &  -- &    \\
NGC 1533  &174&  -- &  -- &      & NGC 6958  &223& 171  & 137 & L98  \\
NGC 1553  &180& 142 & --    &L98,R88&NGC 7007  & 125&  --  &  -- &  \\
NGC 1947  &142& 142 & 142 &BGZ92 & NGC 7079  &155& 125  &  85 & BG97 \\
NGC 2749  &248& 221 & --  &JS89  & NGC 7097  &224& 234  & 196 & C86 \\
NGC 2911  &235&  -- & --  &      & NGC 7135  &231*& 239  & --  & L98 \\
NGC 2962  &197& 168 & 119 & PS00 & NGC 7192  &257& 247  & 266 & CD94  \\
NGC 2974  &220& 170 & 130 & CvM94 & NGC 7332  &136& 127  & 116 & SP97c\\
NGC 3136  &230& 180 &     & KZ00 & NGC 7377  &145&  --  &  -- & \\
 & & & &  & & & & &  \\
NGC 3258  &271& 313 & 264 & KZ00 & IC 1459   &311& 269  & 269 & FI94 \\
NGC 3268  &227& 155 &  -- & KZ00 & IC 2006   &122& --     & --    & CDB93\\
NGC 3489  &129& 116 & 115 & CMP00& IC 3370   &202& 146  & 127 & J87  \\
NGC 3557  &265& 247 & 220 & FI94 & IC 4296   &340& 310  & 320 & S93  \\
NGC 3607  &220& 210 & 195 & CMP00& IC 5063   &160& --   &  -- &     \\
\hline
\end{tabular}}
\label{table4}

\medskip
{\bf Notes}:  the average central value, obtained from the on-line compilation
HYPERCAT (http://www-obs.univ-lyon1.fr/hypercat/), is adopted for
$\sigma_{r_e/8}$. Values at $\sigma_{r_e/4}$ and
$\sigma_{r_e/2}$ velocity dispersion values are obtained from references
quoted in columns 5 and 10. Legend: DO95 = D'Onofrio et al. \cite{Do95};
SP97a = Simien \& Prugniel (\cite{SP97a}); SP97b = Simien \& Prugniel (\cite{SP97b});
PS97 = Prugniel \& Simien  (\cite{PS97}); PS00 = Prugniel \& Simien (\cite{PS00});
PS98 = Prugniel \& Simien (\cite{PS98});
FI94 = Fried \& Illingworth (\cite{FI94}); Z96 = Zeilinger et al. (\cite{Z96});
JS89 = Jedrzejewski \& Schecther (\cite{JS89}); R88 = Rampazzo (\cite{R88});
L98 = Longhetti et al. (\cite{L98b}); BGZ92 = Bertola et al. (\cite{BBZ92a});
CMP00 = Caon et al. (\cite{CMP00}); KZ00 = Koprolin \& Zeilinger (\cite{KZ00});
CDB93 = Carollo et al. (\cite{CDB93}); S83 = Sharples et al.  (\cite{S83});
B94 = Bertin et al.  (\cite{B94}); V87 = Varnas et al.  (\cite{Va87});
C86 = Caldwell et al.  (\cite{Ca86}); CD94 = Carollo \& Danziger (\cite{CD94});
J87 = Jarvis (\cite{J87}); BG97 = Bettoni \& Galletta (\cite{BG97});
S93 = Saglia et al. (\cite{S93}); CvM94 = Cinzano \& van der Marel  (\cite{CvM94}).
\end{table*}
\section{Observations and data-reduction}

\subsection{Observations}

Galaxies were observed during two separate runs (March and
September 1998) at the 1.5m ESO telescope (La Silla), equipped with 
a Boller \& Chivens spectrograph and a UV coated CCD Fa2048L 
(2048$\times$2048) camera (ESO CCD \#39). Details of the observations and
typical seeing conditions during each run are reported in  Table~2. Table~3.
provides a journal of observations i.e. the object identification (column 1,5),
the observing run (columns 2,6), the slit position angle oriented North through
East (columns 3,7) and the total exposure time (columns 4,8). The spectroscopic
slit was oriented along the galaxy major axis for most  observations. He-Ar
calibration frames were taken before and after each exposure to allow an
accurate wavelength calibration to be obtained. 

\subsection{Data reduction}

Pre-reduction, wavelength calibration and sky subtraction were performed using 
the IRAF\footnote{IRAF is distributed by the National Optical Astronomy
Observatories, which are operated by the Association of Universities for Research in
Astronomy Inc., under cooperative agreement with the National Science Foundation}
package. 

A marginal misalignment between CCD pixels and the slit has been checked
and corrected on each image by means of an "ad hoc" written routine.
The wavelength range covered by the observations was $\approx$ 3550 -- 9100
\AA. Fringing seriously affected observations longward of $\approx$ 7300
\AA. After accurate flat-fielding correction we considered the wavelength 
range 3700 - 7250 \AA for further use. Multiple spectra for a given galaxy were co-added. 
Relative flux calibration was obtained using a sequence of spectrophotometric 
standard stars.
Before flux calibration, frames were corrected for atmospheric extinction,
tailored to the ESO La Silla coefficients. The redshift value of each galaxy was directly
measured from the lines of spectra. Spectra were 
finally de-redshifted to the rest frame. A set of representative spectra of the
galaxies in the sample are presented in Figure~\ref{fig2}. The figure also 
shows the similarity of the two sides of the galaxy with respect to the nucleus:
surprisingly after the geometrical and redshift corrections the two sides compare 
within few (2-3) percent, the major deviations due to the asymmetric distribution
of the emission within the galaxies. For some galaxies, namely NGC~1947, NGC~2911, 
NGC~5328 and NGC~6875 there are serious differences between the two sides
with respect to the nucleus since the spectrum is  contaminated by the presence of a 
foreground star (see also next section). 

\subsection{Extraction of apertures and gradients}

We have extracted flux-calibrated spectra along the slit in seven circular
concentric  regions, hereafter "apertures", and in four adjacent regions,
hereafter "gradients". Aperture spectra were sampled with radii of 1.5\arcsec,
2.5\arcsec, 10\arcsec, r$_e$/10, r$_e$/8, r$_e$/4 and r$_e$/2. Our aperture
spectra are suitable for comparison with typical apertures commonly used in the
literature both with different galaxies sample and  with ongoing galaxy surveys
(e.g.  the SLOAN fiber spectra). The
apertures were simulated by assuming that each radial point along the semi-major
axis (sampled at both sides of our slit) is representative of the corresponding
semi-ellipsis in the two-dimensional image. The galaxy  ellipticity,
$\varepsilon$ is given in Table~\ref{table1} col. (13) and has been assumed to be
constant with radius. With this trivial relation between a point within the
simulated semi-circular apertures and the spectrum along the semi-major axis on
the same side, we have calculated the average surface brightness spectrum and the
corresponding  {\it luminosity weighted} radius, $r_l$, of each semi-circular
aperture. The average radius and the flux in each aperture are given by the
formulae:
\begin{equation}
\label{1}
\langle r_l \rangle = \frac{\int r F_{\lambda}(r,\varepsilon) ds}{\int F_{\lambda}(r,\varepsilon) ds}
\end{equation}

\begin{equation}
\label{2}
F_{\lambda}= \frac{\int F_{\lambda}(r,\varepsilon) ds}{\int ds}
\end{equation}

where $r$, $\varepsilon$ and $s$ are the radius along the slit, the
ellipticity and  the area respectively. 
This procedure allow us to obtain a fair estimate of the aperture measurement
corresponding to a mono-dimensional spectrum, in the lack of spectra along the
minor axis (see e.g. Gonz\' alez, 1993 for a thorough discussion).

Simulated aperture spectra sample an increasing concentric
circular region. However, the S/N of our spectra is enough to obtain information
on the spatial gradients. To this purpose we
have extracted spectra also in four adjacent regions along 
each semi-major axis, 0 $\leq$ r $\leq$r$_e$/16 ("nuclear"), 
r$_{e}$/16 $\leq$ r $\leq$r$_e$/8, r$_{e}$/8 $\leq$ r $\leq$r$_e$/4 and
r$_{e}$/4 $\leq$ r $\leq$r$_e$/2) providing the linear average flux
in the above interval. The average radius and the flux in each interval are given
by the formulae:
\begin{equation}
\label{3}
\langle r_l \rangle = \frac{\int r F_{\lambda}(r) dr}{\int
F_{\lambda}(r) dr}
\end{equation}

\begin{equation}
\label{4}
F_{\lambda} = \frac{\int F_{\lambda} dr}{\int dr}
\end{equation}

Figure~\ref{fig2} shows the spectra of the gradient of IC~1459 as a representative 
galaxy. Each panel displays the spectra extracted from
the two symmetric sides with respect to the nucleus. The strategy of averaging the
two sides of the spectrum with respect to the nucleus deserves few comments. 
Each single galaxy demonstrates to be quite homogeneous because in general the
variations of the opposite sides are well within a few (2-3) percent in most
of the cases. Even very faint features are well replicated 
in each side suggesting both that they are real photospheric features and
a radial homogeneity of the stellar population. 

In few cases, we notice that the emission features are less prominent 
(or even absent) in one side of the galaxy with respect to the other.
The ionized gas component is less homogeneously distributed than the
stellar component. This could suggest that the gas is not in an equilibrium 
configuration in the potential well of the galaxy and possibly accreted 
from outside. An alternative explanation could be that the excitation
mechanism is local (see also Appendix A). The study of 
the emission features as function of the distance from the galaxy
center will be deal with in a forthcoming paper.  

Given the homogeneity of the side-spectra we present indices for the averaged spectra.
However,  due to the contamination of the spectrum by foreground stars we measured 
for  NGC~1947, NGC~2911 and NGC~6875 apertures and gradients up to
r$\leq$r$_{e}$/8, while for NGC~5328 we consider apertures and gradients 
up to r$\leq$r$_{e}$/4.

\begin{figure}
\psfig{figure=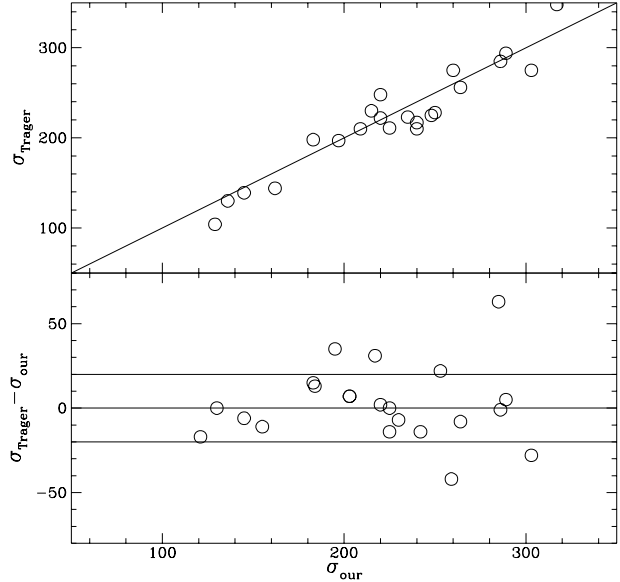,width=9cm,clip=}
\caption{Comparison between central velocity dispersions 
used in this paper and in Trager et al. (\cite{Tra98}). The lines indicate an 
average error (20 km~s$^{-1}$) in the central velocity dispersion measurements.}
\label{fig3}
\end{figure}

\begin{table}
\small{
\begin{tabular}{lcc}
& &   \\
\multicolumn{3}{c}{\bf Table 5. Lick standard stars} \\
\hline\hline
\multicolumn{1}{c}{Ident.}
& \multicolumn{1}{c}{Spectral Type}
& \multicolumn{1}{c}{Numb. obser.s}
 \\
\hline
HD~165195 & K3p     &  1 \\
HD~172401 & K0III   &  1 \\
HD~23430  & A0      &  2 \\ 
HR~6159   & K4III   &  2 \\ 
HR~6710   & F2IV    &  1 \\ 
HR~6770   & G8III   &  1 \\ 
HR~6868   & M0III   &  1 \\ 
HR~7317   & K4III   &  2 \\ 
HR~3145   & K2III   &  3 \\ 
HR~3418   & K1III   &  3 \\ 
HR~3845   & K2.5III &  1 \\ 
HR~4287   & K1III   &  2 \\ 
HR~5480   & G7III   &  1 \\ 
HR~5582   & K3III   &  2 \\ 
HR~5690   & K5III   &  1 \\ 
HR~5888   & G8III   &  1 \\ 
HR~6299   & K2III   &  1 \\ 
\hline
\end{tabular}}
\label{table5}
\end{table}

\section{Measurements of line-strength indices and transformation to the Lick-IDS System}

In the following sub-sections we detail the procedure we adopted to
extract line--strength indices from the original spectra and to transform them
into the Lick--IDS System. We measured  
21 line-strength indices of the original Lick-IDS system using the  redefined
passbands (see Table 2 in Trager et al. \cite{Tra98} for the index definitions) plus 4 
new line strength indices introduced by Worthey \& Ottaviani  (\cite{OW97}) 
(see their Table~1 for the index definitions and Table 2 in Trager et al. \cite{Tra98}). 
In the subsequent analysis we then derived this set of 25 indices.
We tested our index-measuring pipeline on the original Lick
spectra comparing our measurements with the index values given by Worthey {\tt
http://astro.wsu.edu/worthey/html/system.html}.

\subsection{Spectral resolution}

Our spectral resolution (FWHM $\sim$ 7.6 \AA\  at $\sim$ 5000 \AA) on the entire
spectrum  is slightly larger than the wavelength-dependent resolution of the Lick--IDS
system (see Worthey \& Ottaviani \cite{OW97}). In order to conform our measures
to the Lick-IDS System, we  smoothed our data convolving each spectrum
(apertures and gradients) with a wavelength-dependent gaussian kernel with
widths given by the formula:

\begin{equation}
\label{5}
\sigma_{smooth}(\lambda) = \sqrt{\frac{FWHM(\lambda)^2_{Lick} - 
FWHM(\lambda)^2_{our}}{8 \ln 2}}
\end{equation}

The selection of a gaussian kernel is justified by the gaussian shape of both
our and Lick spectra (Worthey \& Ottaviani \cite{OW97}) absorption lines. 

\subsection{Correction for velocity dispersion}

The observed spectrum of a galaxy can be regarded as a stellar spectrum
(reflecting the global spectral characteristics of the galaxy) convolved with
the radial velocity distribution of its stellar population.  Therefore spectral
features in a galactic spectrum are not the simple sum of its corresponding
stellar spectra, because of the stellar motions.  To measure the stellar
composition of galaxies, we need to correct index measurements for the
effects of the galaxy velocity dispersion (see e.g. G93, Trager et al.
(\cite{Tra98}); Longhetti et al. (\cite{L98a})). 

To this purpose, among the Lick stars  observed together with the
galaxies (see also Section~4.4), we have selected stars with spectral type
between G8III and K2III (7 stars in our sample) typically used as
kinematical templates in early-type galaxies.  The list of the observed stars, as
well as their spectral type, is given in Table~5. The
stellar spectra (degraded to the Lick resolution)  have been convolved
with gaussian curves of various widths in order to simulate different
galactic velocity dispersions. We have considered a grid of velocity dispersion
values in the range $(80-350) km s^{-1}$. The values of velocity
dispersion adopted for line-strength correction in 
the present paper are in agreement with those adopted by 
Trager et al. (\cite{Tra98}), as shown in Figure~\ref{fig3}, and well
within the above velocity dispersion range . 
On each convolved spectrum we have measured the 25 Lick-indices. 
The fractional index variations have been derived for each velocity dispersion,
$\sigma$, of our grid through an average on the selected stellar spectra:
 
\begin{equation}
\label{6}
R_{i,\sigma}= \frac{1}{N} \sum_{j=1}^N (\frac{EW_{i,j,\sigma}-EW_{i,j,0}}{EW_{i,j,0}}) 
\end{equation}

where N is the number of studied stars, $EW_{i,j,\sigma}$ is the i-th index
measured on the j-th star at the velocity dispersion $\sigma$, and $EW_{i,j,0}$
is the measured index at zero velocity dispersion. 

To compute the corrections for velocity dispersion, we derived at the radius 
of each aperture and gradient the corresponding $\sigma$ value using the data 
listed in Table~4. The tabulated values characterize the trend of each galaxy 
velocity dispersion curve. 
For galaxies having only the central (r$_e$/8) value of  $\sigma$ we use this value also
for the correction of the indices at larger radii (the tables of indices
uncorrected for velocity dispersion are available,  under request,  to authors 
which may apply suitable corrections when new extended velocity dispersion curve 
measures will be available).

The new index corrected for the effect of velocity dispersion is computed 
in the following way:

\begin{equation}
\label{7}
EW_{i,new}=EW_{i,old}/(1+R_{i,sigma})
\end{equation} 
  
where  $R_{i,\sigma}$ is determined by interpolation of the $\sigma$ value
on the grid of velocity dispersions.

\begin{figure*}
\resizebox{17cm}{!}{ \psfig{figure=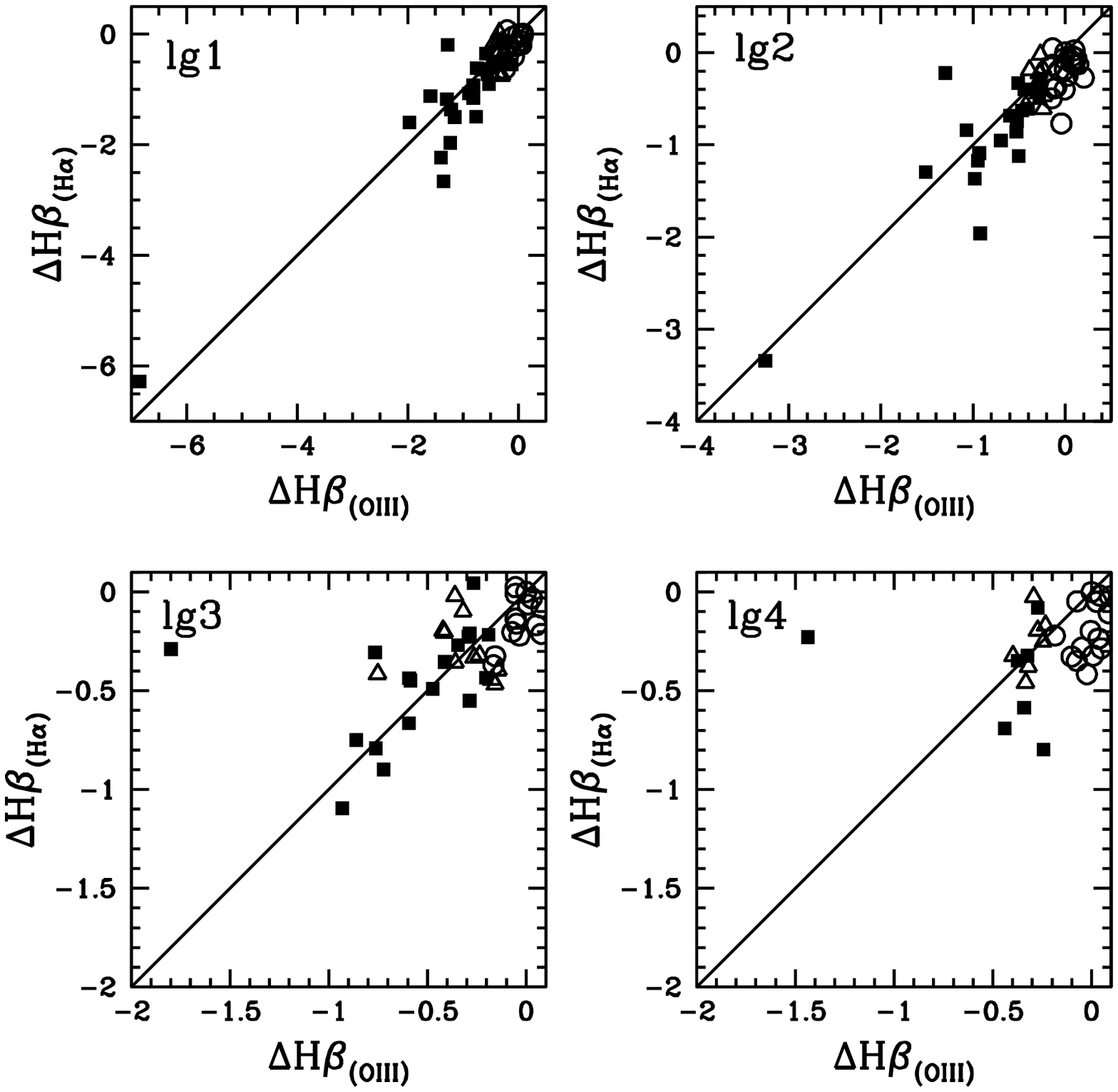,width=16cm,clip=}
\psfig{figure=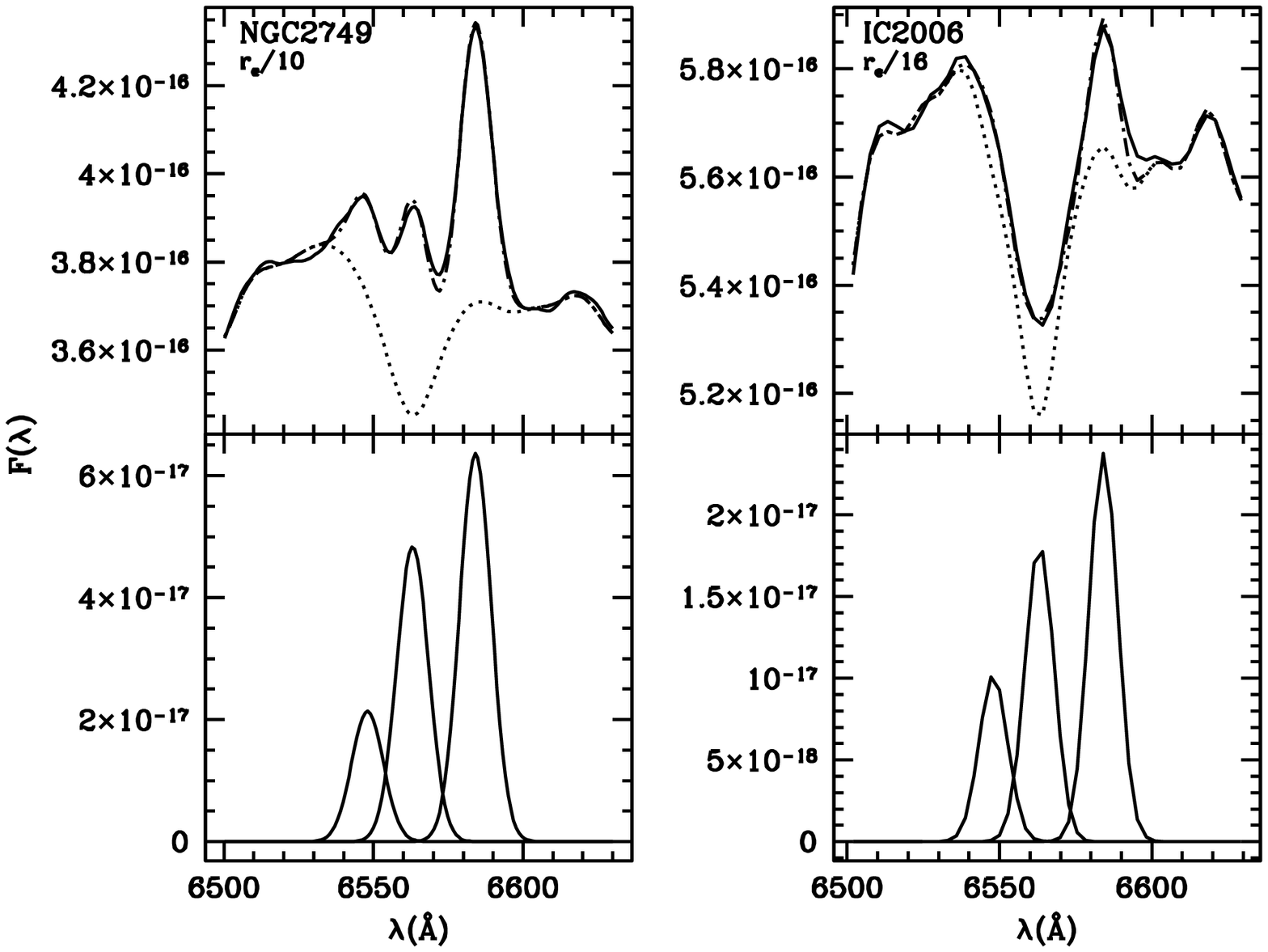,width=14.5cm,clip=} } 
\caption{(left four panels) Comparison between the correction estimate of the  $H\beta$ 
emission derived from $[OIII]$  emission line and from the {H$\alpha$} emission
(see Section 4.3).  The solid line is the one-to-one relation assuming for 
O[III] valid the formula $EW(H\beta_{em})/EW([OIII]\lambda 5007) = 0.7$.
The comparison is shown in the four regions sampled by
the linear gradients (legend: lg1 (0 $\leq$ r $\leq$r$_e$/16("nuclear")), lg2
(r$_{e}$/16 $\leq$ r $\leq$r$_e$/8),  lg3 (r$_{e}$/8 $\leq$ r $\leq$r$_e$/4) and
lg4 (r$_{e}$/4 $\leq$ r $\leq$r$_e$/2). Open circles indicate galaxies which
O[III] emission is detected under 1$\sigma$ level, triangles and full squares
between 1 and 2$\sigma$ levels and above 2$\sigma$level respectively (see text). 
(right four panels) Fitting of N[II]($\lambda$ 6548, 6584)
and H$\alpha$ lines for two representative galaxies: one with H$\alpha$ in emission 
(NGC~2749) and the second with the H$\alpha$ infilling (IC~2006). 
Lower panels show the residuals lines after the subtraction 
of the H$\alpha$ line of the template galaxy NGC~1426 
(dotted lines in the upper panels).}
\label{fig4}
\end{figure*}

\subsection{Correction of the H$\beta$ index  for emission}

The presence of emission lines affects the measure of some line--strength
indices. In particular, the H$\beta$ index measure of the underlying stellar
population could be contaminated by a significant infilling due to presence of
the H$\beta$ emission component.

Gonz\' alez et al. (\cite{G93}) verified a strong correlation 
between the $H\beta$ and the $[OIII]$ emission  in his sample, such that
$EW(H\beta_{em})/EW([OIII]\lambda 5007) = 0.7$. Trager et al. (\cite{Tra00})
examined the accuracy of this correlation by studying the $H\beta /[OIII]$ ratio
supplementing the G93 sample with an additional  sample of early-type galaxies
with emission lines from the catalog of Ho, Filipenko \& Sargent (1997). They
found that the $H\beta /[OIII]$ ratio varies from 0.33 to 1.25, with a median value
of 0.6. They propose that the correction to the $H\beta$ index is
$\Delta$ H$\beta$ = 0.6 EW([OIII]$\lambda$ 5007).

\begin{table}
\small{
\begin{tabular}{lcccc}
& &  & &\\
\multicolumn{5}{c}{\bf Table 6. ~~~  H$\beta$ corrections for apertures} \\
\hline\hline
\multicolumn{1}{c}{Galaxy}
&\multicolumn{1}{c}{aperture}
& \multicolumn{1}{c}{EWO[III]}
& \multicolumn{1}{c}{Quality}
& \multicolumn{1}{c}{EWH$\alpha$}
\\
 \hline 
    NGC128 & 0 &  -0.263 &   2.000 &  -0.558  \\
    NGC128 & 1 &  -0.210 &   1.000 &  -0.532  \\
    NGC128 & 2 &  -0.205 &   1.000 &  -0.481 \\  
    NGC128 & 3 &  -0.182 &   1.000 &  -0.465 \\  
    NGC128 & 4 &  -0.053 &   0.000 &  -0.325 \\ 
    NGC128 & 5 &   0.006 &   0.000 &  -0.301 \\  
    NGC128 & 6 &   0.036 &   0.000 &  -0.186 \\ 
    NGC777 & 0 &   0.039 &   0.000 &  -0.126 \\ 
    NGC777 & 1 &   0.077 &   0.000 &  -0.114 \\
    NGC777 & 2 &   0.060 &   0.000 &  -0.100 \\ 
    NGC777 & 3 &   0.076 &   0.000 &  -0.103 \\  
    NGC777 & 4 &   0.112 &   0.000 &  -0.113 \\ 
    NGC777 & 5 &   0.144 &   0.000 &  -0.124 \\  
    NGC777 & 6 &   0.071 &   0.000 &  -0.124 \\  
    ........       & .....&   ......   &  ......     &  .......  \\
\hline
\end{tabular}}
\label{table6}

\medskip
The table provides in column 2 the number of the aperture, the correspondent radius of which 
is given in Table~9, and the estimate of the EW of the O[III] and H$\alpha$ (columns 3 and 5
respectively) obtained from the subtraction of  the template galaxy, NGC 1426. 
Column 4 gives the quality of the measure, obtained as the ratio between the
estimated emission (F$_{gal}$ - F$_{templ}$) and the variance
of the spectrum in the O[III] wavelength region, $\sigma_\lambda$. 
When the emission is lower than the variance the quality is set to 0. 
When the emission is between 1 and 2  $\sigma_\lambda$ or larger than 2  $\sigma_\lambda$
the quality is set to 1 and 2 respectively. The entire table is given is electronic form.
\end{table}

\begin{table}
\small{
\begin{tabular}{lcccc}
& &  & &\\
\multicolumn{5}{c}{\bf Table 7. ~~~  H$\beta$ corrections for gradients} \\
\hline\hline
\multicolumn{1}{c}{Galaxy}
&\multicolumn{1}{c}{aperture}
& \multicolumn{1}{c}{EWO[III]}
& \multicolumn{1}{c}{Quality}
& \multicolumn{1}{c}{EWH$\alpha$}
\\
 \hline         
    NGC128 & 0 &  -0.317 &   1.000 &  -0.702 \\  
    NGC128  &1 &  -0.204 &   1.000 &  -0.608 \\  
    NGC128 & 2 &  -0.173 &   2.000 &  -0.434 \\  
    NGC128  & 3 &  0.119 &   0.000 &  -0.174  \\
    NGC777  & 0 &   0.042 &  0.000 &  -0.124  \\
    NGC777  & 1 &   0.057 &   0.000 &  -0.106 \\  
    NGC777  & 2 &   0.045 &   0.000 &  -0.168 \\  
    NGC777  & 3 &   0.162 &   0.000 &  -0.129 \\ 
     ....           & ... &   .....     &   .....     &  .....  \\
\hline
\end{tabular}}
\label{table7}

\medskip
The table provides in column 2 the gradient, 
(0=0 $\leq$ r $\leq$r$_e$/16, 1 = r$_{e}$/16 $\leq$ r $\leq$r$_e$/8, 
2 = r$_{e}$/8 $\leq$  r $\leq$r$_e$/4 and 3 = r$_{e}$/4 $\leq$ r $\leq$r$_e$/2), 
and the estimate of the EW of the O[III] and H$\alpha$ (columns 3 and 5
respectively) obtained from the subtraction of  the template galaxy, NGC 1426. 
Column 4 gives the quality of the measure, obtained as the ratio between the
estimated emission (F$_{gal}$ - F$_{templ}$) and the variance
of the spectrum in the O[III] wavelength region, $\sigma_\lambda$. 
When the emission is lower than the variance the quality is set to 0. 
When the emission is between 1 and 2  $\sigma_\lambda$ or larger than 2  $\sigma_\lambda$
the quality is set to 1 and 2 respectively. The entire table is given in electronic form.
\end{table}

The first step in order to measure the EW([OIII]$\lambda 5007$)  of the emission line 
is to degrade  each spectrum (apertures and gradients)   to the Lick resolution.
The second, more delicate step, is to ``build'' a suitable template for the underlying
stellar component and then adapt it to  the galaxy velocity dispersion. 
At this purpose different methods have been adopted both using stellar 
and galaxy templates.

Gonz\' alez~(\cite{G93}) used stellar templates.  The adopted technique adopted 
consists in simultaneously fitting  
the kinematics and the spectrum of a galaxy with a library of stellar spectra. 
However, we are aware that absorption line spectra of early-type
galaxies cannot be adequately fitted using  Galactic stars or star clusters, the
main reason being the high metallicity in giant ellipticals and the non-solar
element ratios in ellipticals. To compute the emissions Goudfrooij (\cite{Gou98}) 
suggested to use a suitable template galaxy spectrum of an elliptical. Following his
suggestion we then considered galaxies in our sample that, according to our observed 
spectrum and the combined information coming from the literature, show neither
evidence in their spectrum of neither emission features   nor of dust,
usually associated with the gas emission (see Goudfrooij \cite {Gou98}),
in their image. To this purpose we adopted  NGC~1426 spectrum as a template
for an old population: this choice is motivated both by the lack of
emission line in its spectrum and by the absence of dust features in
high resolution HST images obtained  by  Quillen et
al.~(\cite{Qui00}) (see Appendix A). We then proceeded in following way: 
we smoothed the template spectrum to adapt it to the velocity dispersion
of the galaxy region under exam and normalized it to the galaxy continuum 
on both sides  of H$\alpha$ line. All spectra (aperture and gradients) have been analyzed
using the template in the corresponding region.
NGC 1426 has a low value of velocity dispersion consistently with its low Mg$_2$ index;
this indicates that it is not a giant elliptical and may be representative of the
metal poor tail in our sample. Given the anti-correlation between H$\beta$ index
and Mg$_2$ index strength one may wonder whether this galaxy is the suitable template
for all the sample. In order to check the reliability of the use of this
template we have compared the H$\alpha$ absorption profile of NGC~1426 
with that of  NGC 1407, which belongs to upper tail of the Mg$_2$ $\sigma$
relation. Once adopted the above procedure  of smoothing and normalization
we notice that the residual difference in the H$\alpha$ profile implies a negligible
difference in the computed EW H$\beta$ correction ($\simeq$ 0.03 \AA).
  
We characterized the emission as the flux in excess with respect to the template
within the bandpass 
(4996.85 - 5016.85) centered at 5007\AA, while the continuum is defined by a blue
(4885.00 - 4935.00) and a red (5030.00 - 5070.00) bandpass (Gonz\' alez~\cite{G93}):

\begin{equation}
\label{8} 
EW_{em}=\int_{\lambda_1}^{\lambda_2} \frac{F_R-F_{temp}}{F_C} d \lambda
\end{equation}

where $F_R$, $F_{temp}$ and $F_C$ are  the galaxy, the template and the continuum 
fluxes respectively. According to this definition, detected emissions result as negative EWs. 
Considering the ([OIII]$\lambda 5007$) emissions detected above 1 $\sigma$
(the variance of the spectrum), we derived the EW of the H$\beta$ emission 
from the equation $\Delta$ H$\beta$ = 0.7 EW([OIII]$\lambda$ 5007). 
The derived corrections for H$\beta$ could be easily compared with Gonz\' alez~(\cite{G93}) 
for the three galaxies in common (namely NGC~ 1453,  NGC~4552 and NGC~5846).
We obtained an 0.48 (vs. 0.89$\pm$0.06), 0.02 (vs. 0.25$\pm$0.05) and 0.09 (vs. 0.39$\pm$0.08) 
i.e. systematically smaller corrections as if our template had a residual
gas infilling, but which is not confirmed by imaging observations as 
outlined above. We tested also the use of a stellar template taken from our 
observed  Lick stars, the main problem being that our set of stars is very ``limited" 
with respect to that of Lick stellar library used by Gonz\' alez. The use of 
stellar templates (K giant stars) in our sample, maintaining the above 
O[III] bandpasses,  implies a worse match of the spectral features in our 
galaxy sample than if we adopt NGC~1426 as a template.  Finally this 
results in a systematically higher H$\beta$ corrections,
at least for the galaxies in common with Gonz\' alez.

The large wavelength coverage of our spectra permits us to measure also 
the H$\alpha$ emission  and allows a further estimate of the H$\beta$ emission
according to the relation  $F_{H\beta} =  1/2.86 F_{H\alpha}$ (see e.g. 
Osterbrock~\cite{Osterb89}).
  
The measure of the $H\alpha$ emission is not straightforward
in our spectra since the line is  blended with the ([NII]$\lambda 6548,
6584$) emission lines. To derive the H$\alpha$ emissions we fitted
 each galaxy spectrum (apertures and gradients)  with a model resulting from
 the sum of our template galaxy spectrum and three gaussian curves of arbitrary
 widths and amplitudes (see Figure~\ref{fig4}). Once derived the H$\beta$ emitted 
fluxes from  equation~(\ref{8}) we computed the pseudo-continua in H$\beta$ according
 to the bandpass definition of Trager et al. (\cite{Tra98}) and used them to transform flux measures
 into EWs. 

In Figure~\ref{fig4} (left panels) we plot the comparison  between the two different 
estimates computed in the four gradients. The points in the figure do not include 
the Seyfert 2 galaxy IC~5063 for which the $\Delta H\beta$ as derived from 
the [OIII] emission is significantly higher than the value resulting 
from $H\alpha$ emission.

The new H$\beta$ index, corrected for the emission infilling, is computed 
from the non raw one according to the formula 
$EW (H\beta_{corr}) = EW(H\beta_{raw}) - H \beta_{em}$, where EW(H$\beta_{corr}$) 
is the corrected value obtained applying the $H \beta_{em}$ estimate derived 
from the [OIII] emission  using as  template the galaxy NGC~1426. 
This latter estimate is {\it statistically similar} to that 
obtained from the  H$\alpha$ correction as shown in Figure~\ref{fig4}, although
the use of H$\alpha$ estimate for emission correction will be widely discussed 
in a forthcoming paper.

Table~6 and Table~7 report the values of the   H$\beta$ correction for the
apertures and gradients derived from EWO[III] and H$\alpha$ (complete tables are 
given in electronic form).    

\begin{table}
\small{
\begin{tabular}{lcccc}
& &  & & \\
\multicolumn{5}{c}{\bf Table 8 ~~~~~ $\alpha$ and $\beta$ coeff. for index correction} \\
\hline\hline
\multicolumn{1}{c}{Index}
& \multicolumn{1}{c}{$\alpha$}
& \multicolumn{1}{c}{$\beta$}
& \multicolumn{1}{c}{aver. disp.}
& \multicolumn{1}{c}{unit}
 \\
 & &  & &\\
\hline
CN$_1$           &         1.059  &         0.023  & 0.025   & mag\\
CN$_2$           &         1.035  &         0.030  & 0.023   & mag \\
Ca4227        &         1.317  &         0.408  & 0.396   & \AA \\
G4300         &         1.105  &         0.179 &  0.310  & \AA\\
Fe4383        &         0.963  &         1.169  & 0.772   & \AA \\
ca4455        &         0.451  &         1.844  & 0.341   & \AA \\
Fe4531        &         1.289  &        -0.299  &0.437    & \AA\\
Fe4668        &         0.976  &         0.128  &0.653    & \AA \\
H$\beta$    &         1.064  &        -0.196   &0.166    & \AA \\
Fe5015        &         1.031  &         0.804  & 0.396   & \AA \\
Mg$_1$           &         1.014  &         0.015  &  0.009  & mag \\
Mg$_2$           &         0.998  &         0.020   & 0.012  &  mag \\
Mgb           &         1.014  &         0.417  &  0.241  & \AA \\
Fe5270        &         1.058  &         0.270  & 0.240   & \AA\\
Fe5335        &         0.990  &         0.356  & 0.249   & \AA \\
Fe5406        &         1.005  &         0.282  & 0.151   & \AA \\
Fe5709        &         1.321  &        -0.270  & 0.174   & \AA \\
Fe5782        &         1.167  &        -0.075 &  0.165  & \AA \\
NaD           &          1.003  &         0.027  & 0.245   & \AA \\
TiO$_1$          &          0.997  &         0.004  & 0.006   &  mag\\
TiO$_2$          &          1.003  &        -0.001   &0.008    &   mag \\
H$\delta_{A}$ &         1.136  &        -0.622   &1.087    & \AA \\
H$\gamma_{A}$ &        0.990  &         0.518   & 0.734   & \AA \\
H$\delta_{F}$ &         1.059  &        -0.036   & 0.503   & \AA \\
H$\gamma_{F}$ &         1.011  &         0.458   & 0.745   & \AA \\
\hline
\end{tabular}}
\label{table8}
\end{table}

\begin{figure*}
\psfig{figure=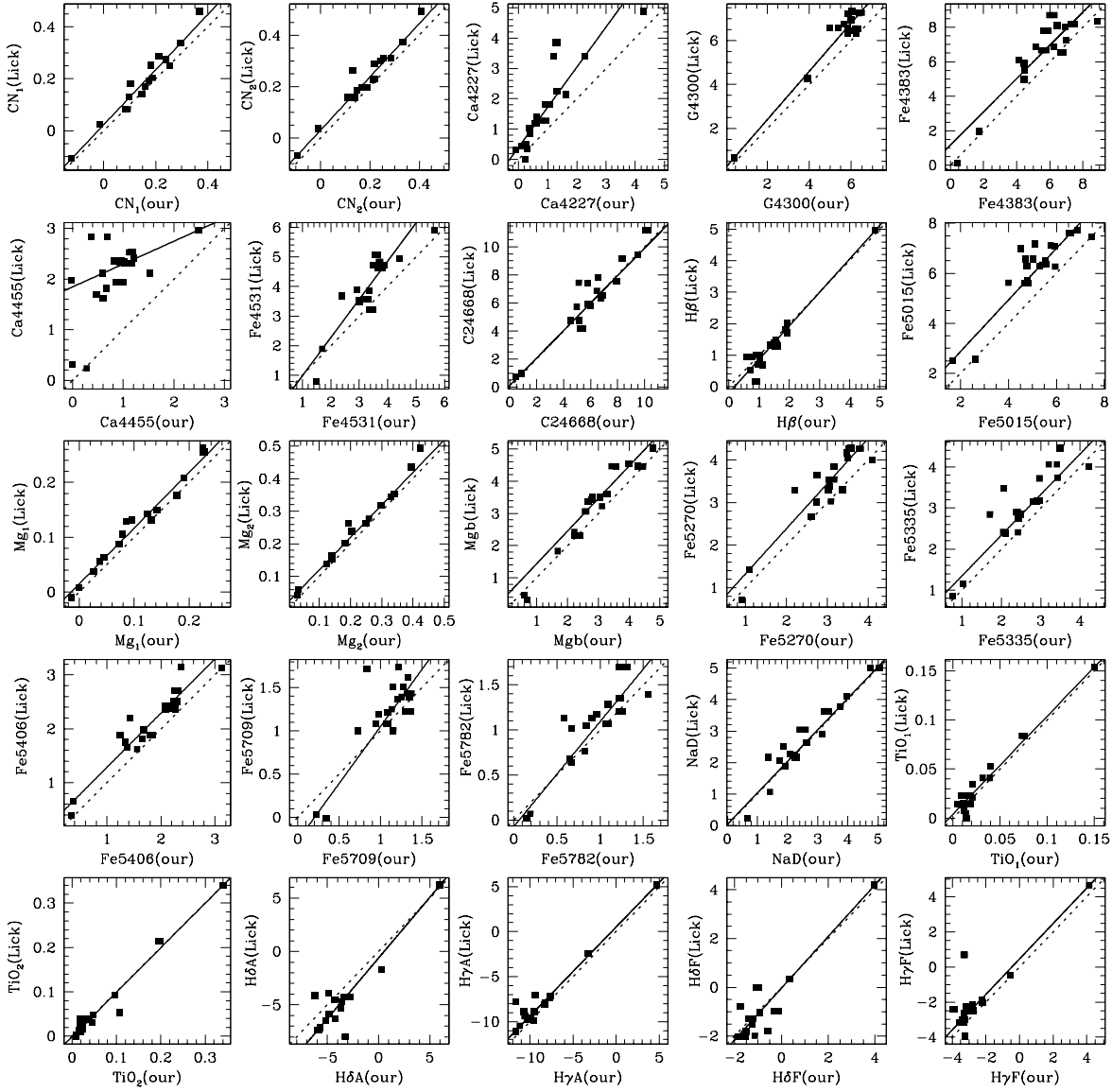,width=16cm,clip=}
\caption{Comparison of passband measurements from our spectra and original
Lick data for 17 Lick standard stars. The dotted line is the one to 
one relation while the solid line is the {\it robust straight-line fit} to the filled squares.} 
\label{fig5}
\end{figure*}

\begin{table*}
\tiny{
\begin{tabular}{lrrrrrrrrrrrrrrrr}
 & & & & & & & & & & & & & & &   \\
\multicolumn{16}{c}{\bf Table 9 ~~~~ Fully corrected line--strength indices for the apertures  } \\
\hline\hline
\multicolumn{1}{c}{galaxy} &
\multicolumn{1}{c}{iz} &
\multicolumn{1}{c}{ie} &
\multicolumn{1}{c}{r$_1$} &
\multicolumn{1}{c}{r$_2$} &
\multicolumn{1}{c}{CN$_1$} &
\multicolumn{1}{c}{CN$_2$} &
\multicolumn{1}{c}{Ca4227} &
\multicolumn{1}{c}{G4300} &
\multicolumn{1}{c}{Fe4383} &
\multicolumn{1}{c}{Ca4455} &
\multicolumn{1}{c}{Fe4531} &
\multicolumn{1}{c}{Fe4668} &
\multicolumn{1}{c}{H$\beta$} &
\multicolumn{1}{c}{Fe5015}&
\multicolumn{1}{c}{other}  \\
\multicolumn{1}{c}{galaxy} &
\multicolumn{1}{c}{iz} &
\multicolumn{1}{c}{ie} &
\multicolumn{1}{c}{r$_l$} &
\multicolumn{1}{c}{r$_e$} &
\multicolumn{1}{c}{eCN$_1$} &
\multicolumn{1}{c}{eCN$_2$} &
\multicolumn{1}{c}{eCa4227} &
\multicolumn{1}{c}{eG4300} &
\multicolumn{1}{c}{eFe4383} &
\multicolumn{1}{c}{eCa4455} &
\multicolumn{1}{c}{eFe4531} &
\multicolumn{1}{c}{eFe4668} &
\multicolumn{1}{c}{eH$\beta$}&
\multicolumn{1}{c}{eFe5015}&
\multicolumn{1}{c}{indices} \\
\hline
    NGC128 &  0 & 0 &  0.000  &   0.087  & 0.136  &   0.167 &  1.830 &    6.150 & 5.790  &   2.330 &    4.150  & 7.580 &    1.460 &    5.860 &  ...  \\  
    NGC128 &  0 & 1 &  0.055  &  17.300  & 0.003  &   0.004 &  0.054 &    0.087 & 0.149  &   0.079 &    0.139  & 0.225 &    0.099 &    0.198 & ... \\    
    NGC128 &  1 & 0 &  0.000  &   0.100  & 0.136  &   0.166 &  1.830 &    6.150 & 5.850  &   2.340 &    4.150  & 7.610 &    1.410 &    5.880 &    ...  \\ 
    NGC128 &  1 & 1 &  0.062  &  17.300  & 0.003  &   0.004 &  0.054 &    0.087 & 0.149  &   0.080 &    0.139  & 0.226 &    0.098 &    0.198  &   ...  \\   
    NGC128 &  2 & 0 &  0.000  &   0.125  & 0.135  &   0.165 &  1.830 &    6.150 & 5.950  &   2.350 &    4.160  & 7.660 &    1.440 &    5.910 &  ...  \\  
    NGC128 &  2 & 1 &  0.076  &  17.300  & 0.003  &   0.004 &  0.054 &    0.087 & 0.150  &   0.080 &    0.141  & 0.223 &    0.099 &    0.198 &    ... \\  
    NGC128 &  3 & 0 &  0.000  &   0.145  & 0.135  &   0.164 &  1.840 &    6.150 & 5.980  &   2.350 &    4.150  & 7.680 &    1.420 &    5.950  &... \\   
    NGC128 &  3 & 1 &  0.087  &  17.300  & 0.003  &   0.004 &  0.055 &    0.087 & 0.151  &   0.080 &    0.141  & 0.224 &    0.099 &    0.199 &   ... \\
    NGC128 &  4 & 0 &  0.000  &   0.250  & 0.132  &   0.159 &  1.760 &    6.120 & 5.910  &   2.300 &    4.090  & 7.630 &    1.310 &    5.970 &  ... \\   
    NGC128 &  4 & 1 &  0.142  &  17.300  & 0.003  &   0.004 &  0.056 &    0.090 & 0.154  &   0.082 &    0.144  & 0.225 &    0.098 &    0.206 &  ... \\  
    NGC128 &  5 & 0 &  0.000  &   0.289  & 0.131  &   0.158 &  1.770 &    6.120 & 5.920  &   2.300 &    4.100  & 7.640 &    1.300 &    6.000 & ... \\  
    NGC128 &  5 & 1 &  0.161  &  17.300  & 0.003  &   0.004 &  0.056 &    0.091 & 0.155  &   0.083 &    0.145  & 0.228 &    0.099 &    0.207 &  ... \\  
    NGC128 &  6 & 0 &  0.000  &   0.500  & 0.125  &   0.151 &  1.760 &    6.090 & 5.970  &   2.300 &    4.130  & 7.670 &    1.340 &    6.090 &   ... \\  
    NGC128 &  6 & 1 &  0.260  &  17.300  & 0.003  &   0.004 &  0.059 &    0.095 & 0.160  &   0.086 &    0.152  & 0.237 &    0.103 &    0.216 &   ... \\  
    NGC777 &  0 & 0 &  0.000  &   0.044  & 0.176  &   0.215 &  2.410 &    6.030 & 5.050  &   2.290 &    3.910  & 8.820 &    1.150 &    6.220 & ... \\  
    NGC777 &  0 & 1 &  0.028  &  34.400  & 0.002  &   0.003 &  0.046 &    0.083 & 0.131  &   0.070 &    0.122  & 0.181 &    0.079 &    0.176 &  ... \\  
    NGC777 &  1 & 0 &  0.000  &   0.073  & 0.175  &   0.212 &  2.050 &    6.100 & 4.940  &   2.290 &    4.030  & 8.450 &    1.200 &    6.260 &  ... \\   
    NGC777 &  1 & 1 &  0.045  &  34.400  & 0.004  &   0.004 &  0.055 &    0.104 & 0.174  &   0.090 &    0.163  & 0.243 &    0.107 &    0.227 & ... \\  
    NGC777 &  2 & 0 &  0.000  &   0.100  & 0.173  &   0.209 &  1.870 &    6.120 & 4.980  &   2.340 &    4.050  & 8.360 &    1.210 &    6.190  &   ...\\   
    NGC777 &  2 & 1 &  0.059  &  34.400  & 0.004  &   0.004 &  0.053 &    0.101 & 0.168  &   0.086 &    0.158  & 0.235 &    0.104 &    0.220 &   ...\\  
    NGC777 &  3 & 0 &  0.000  &   0.125  & 0.167  &   0.205 &  1.870 &    6.160 & 5.140  &   2.400 &    3.990  & 8.240 &    1.270 &    6.170 &   ...  \\  
    NGC777 &  3 & 1 &  0.072  &  34.400  & 0.002  &   0.003 &  0.046 &    0.080 & 0.127  &   0.065 &    0.115  & 0.174 &    0.075 &    0.165 &   ...\\   
    NGC777 &  4 & 0 &  0.000  &   0.145  & 0.165  &   0.203 &  1.870 &    6.160 & 5.160  &   2.440 &    3.940  & 8.150 &    1.300 &    6.200 &   ... \\   
    NGC777 &  4 & 1 &  0.081  &  34.400  & 0.002  &   0.003 &  0.045 &    0.080 & 0.127  &   0.065 &    0.115  & 0.175 &    0.075 &    0.165 &...\\  
    NGC777 &  5 & 0 &  0.000  &   0.250  & 0.159  &   0.194 &  1.860 &    6.250 & 5.070  &   2.470 &    3.820  & 8.220 &    1.190 &    6.230 &   ...\\  
    NGC777 &  5 & 1 &  0.125  &  34.400  & 0.002  &   0.003 &  0.046 &    0.077 & 0.123  &   0.064 &    0.115  & 0.169 &    0.073 &    0.164  &  ... \\  
    NGC777 &  6 & 0 &  0.000  &   0.500  & 0.174  &   0.199 &  1.920 &    6.440 & 5.640  &   2.530 &    3.960  & 7.910 &    1.400 &    6.310 &  ... \\  
    NGC777 &  6 & 1 &  0.210  &  34.400  & 0.005  &   0.005 &  0.074 &    0.130 & 0.216  &   0.108 &    0.204  & 0.298 &    0.131 &    0.293 &  ...\\  
    ...  & ... & ... & ... & ... & ... & ... & ... &... & ... & ... & ... & ... & ... & ... & ... \\
\hline
\label{table9}
\end{tabular}}

\medskip{The complete Table~9 is given in electronic form.} 
\end{table*}

\begin{table*}
\tiny{
\begin{tabular}{lrrrrrrrrrrrrrrrr}
 & & & & & & & & & & & & & & &  \\
\multicolumn{16}{c}{\bf Table 10 ~~~~ Fully corrected line--strength indices for the gradients} \\
\hline\hline
\multicolumn{1}{c}{galaxy} &
\multicolumn{1}{c}{iz} &
\multicolumn{1}{c}{ie} &
\multicolumn{1}{c}{r$_1$} &
\multicolumn{1}{c}{r$_2$} &
\multicolumn{1}{c}{CN$_1$} &
\multicolumn{1}{c}{CN$_2$} &
\multicolumn{1}{c}{Ca4227} &
\multicolumn{1}{c}{G4300} &
\multicolumn{1}{c}{Fe4383} &
\multicolumn{1}{c}{Ca4455} &
\multicolumn{1}{c}{Fe4531} &
\multicolumn{1}{c}{Fe4668} &
\multicolumn{1}{c}{H$\beta$} &
\multicolumn{1}{c}{Fe5015}&
\multicolumn{1}{c}{other}  \\
\multicolumn{1}{c}{galaxy} &
\multicolumn{1}{c}{iz} &
\multicolumn{1}{c}{ie} &
\multicolumn{1}{c}{r$_l$} &
\multicolumn{1}{c}{r$_e$} &
\multicolumn{1}{c}{eCN$_1$} &
\multicolumn{1}{c}{eCN$_2$} &
\multicolumn{1}{c}{eCa4227} &
\multicolumn{1}{c}{eG4300} &
\multicolumn{1}{c}{eFe4383} &
\multicolumn{1}{c}{eCa4455} &
\multicolumn{1}{c}{eFe4531} &
\multicolumn{1}{c}{eFe4668} &
\multicolumn{1}{c}{eH$\beta$}&
\multicolumn{1}{c}{eFe5015}&
\multicolumn{1}{c}{indices} \\
\hline
    NGC128 & 0 & 0  &   0.000  &  0.063 &     0.140 &     0.167 &    1.580 &    6.020 &    4.840 &    2.160 &    3.820 &    7.080 &    1.540 &    5.520 &  ...  \\
    NGC128 & 0 & 1  &   0.031  & 17.300 &     0.004 &     0.005 &    0.077 &    0.132 &    0.219 &    0.112 &    0.191 &    0.318 &    0.127 &    0.262 &  ... \\
    NGC128 & 1 & 0  &   0.063  &  0.125 &     0.137 &    0.168  &   1.790  &   6.150  &   5.680  &   2.290  &   4.210  &   7.570  &   1.340  &   5.830  & ...  \\
    NGC128 & 1 & 1  &   0.093  &  17.300&     0.003 &    0.004  &   0.053  &   0.094  &   0.149  &   0.079  &   0.157  &   0.234  &   0.104  &   0.200  &  ...  \\
    NGC128 & 2 & 0  &   0.125  &   0.250&     0.130 &    0.160  &   1.800  &   6.160  &   6.060  &   2.330  &   4.010  &   7.740  &   1.560  &   5.940  & ...  \\
    NGC128 & 2 & 1  &   0.182  &  17.300&     0.003 &    0.004  &   0.054  &   0.094  &   0.148  &   0.080  &   0.153  &   0.237  &   0.103  &   0.199  &  ...  \\
    NGC128 & 3 & 0  &   0.250  &   0.500&     0.138 &    0.162  &   1.850  &   6.030  &   6.110  &   2.340  &   4.130  &   7.630  &   1.330  &   6.330  & ...  \\
    NGC128 & 3 & 1  &   0.359  &  17.300&     0.003 &    0.004  &   0.057  &   0.101  &   0.158  &   0.084  &   0.163  &   0.252  &   0.108  &   0.211  &  ...  \\
    NGC777 & 0 & 0  &   0.000  &   0.063&     0.175 &    0.213  &   2.280  &   5.980  &   4.990  &   2.270  &   3.970  &   8.820  &   1.210  &   6.140  & ...  \\
    NGC777 & 0 & 1  &   0.029  &  34.400&     0.003 &    0.004  &   0.063  &   0.114  &   0.194  &   0.094  &   0.163  &   0.267  &   0.104  &   0.233  &  ...  \\
    NGC777 & 1 & 0  &   0.063  &   0.125&     0.170 &    0.206  &   1.670  &   6.170  &   5.130  &   2.450  &   4.100  &   8.130  &   1.210  &   6.180  & ...  \\
    NGC777 & 1 & 1  &   0.090  &  34.400&     0.003 &    0.003  &   0.047  &   0.083  &   0.136  &   0.066  &   0.117  &   0.185  &   0.076  &   0.184  &  ...  \\
    NGC777 & 2 & 0  &   0.125  &   0.250&     0.146 &    0.180  &   1.760  &   6.250  &   5.000  &   2.530  &   3.700  &   8.120  &   1.230  &   6.330  & ...  \\
    NGC777 & 2 & 1  &   0.177  &  34.400&     0.003 &    0.004  &   0.056  &   0.094  &   0.155  &   0.078  &   0.139  &   0.211  &   0.095  &   0.212  &  ...  \\
    NGC777 & 3 & 0  &   0.250  &   0.500&     0.189 &    0.210  &   2.040  &   7.080  &   5.970  &   2.430  &   4.060  &   7.880  &   1.830  &   5.970  & ...  \\
    NGC777 & 3 & 1  &   0.347  &  34.400&     0.004 &    0.005  &   0.076  &   0.131  &   0.218  &   0.108  &   0.193  &   0.299  &   0.123  &   0.302  &  ...  \\
     ... & ... & ... & ... & ... & ... & ... &... & ... & ... & ... & ... & ... & ... & ...  & ...  \\
\hline
\label{table10}
\end{tabular}}

\medskip{The complete Table~10 is given in electronic form.} 
\end{table*}

\subsection{Lick-IDS Standard Stars}

After the indices have been homogenized to the Lick-IDS wavelength dependent 
resolution, corrected for emission and velocity dispersion, we still have to perform a last step 
to transform our line-strengths indices into the Lick system. To this purpose we
followed the prescription given by Worthey \& Ottaviani (\cite{OW97}) and
observed, contemporary to the galaxies, a sample of  17 standard stars of
different spectral type common to the Lick library. The observed stars, together
with the corresponding number of observations and the spectral type, 
are given in Table~5. Once the stellar spectra have been degraded to
the wavelength-dependent resolution of the Lick system, we have measured the
line-strength indices with the same procedure adopted for the galaxy spectra. 
We compared our measures with the Lick-IDS indices reported by Worthey et al.
\cite{Wor94} for the standard stars. The deviations of our measures from the Lick
system are parametrized through  a {\it robust straight-line fit}  (see e.g. Numerical
Recipes 1992) which avoid an undesired sensitivity to outlying points in  two
dimension fitting to a straight line. The functional form is $EW_{Lick} = \beta + \alpha \times 
EW_{our}$ where $EW_{our}$  and $EW_{Lick}$ are respectively  
our index measure on the stellar spectrum and the Lick value given in 
Worthey et al. \cite{Wor94}.  

Fig. \ref{fig5} shows the comparison between the Lick indices and our measures 
for the observed standard stars. The dotted line represent the one to one
relation while the solid line is the derived fit. For each index we report the 
coefficients $\alpha$ and $\beta $ of the fit in Table~8. Notice that for the majority 
of the indices $\alpha$ value is very close to 1 and only a zero-point correction is required
(see also Puzia et al.  (\cite{Puzia2002}), although serious deviations from the one-to-one
relation are shown by Ca lines 4227 and 4455 and Fe 4531.

\subsection{Estimate of indices measurement errors}

In order to obtain the errors on each measured index we have used the following
procedure. Starting from a given extracted spectrum (aperture or gradient at different
galactocentric distances), we have generated a  set of
1000 Monte Carlo random modifications, by adding a wavelength dependent 
Poissonian fluctuation from the corresponding spectral noise, $\sigma(\lambda)$.
Then, for each spectrum, we have estimated the moments of the distributions of
the 1000 different realizations of its indices.

\begin{figure*}
\psfig{figure=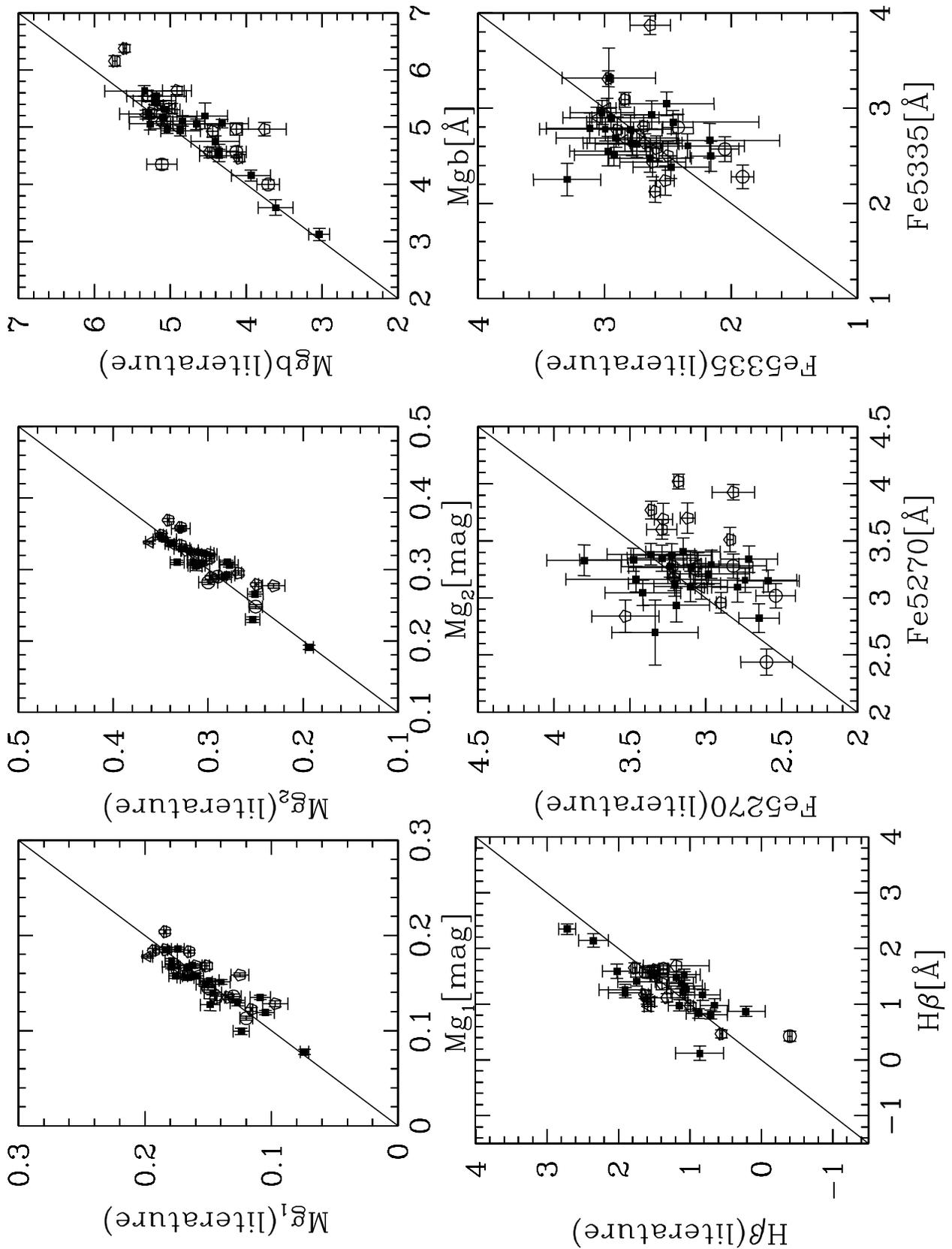,width=\textwidth,angle=0,clip=}
\caption{Comparison of index measurements of Gonz\' alez (1993: open
triangles), Trager (1998: full squares), Longhetti et al. (1998: open circles),  
Beuing et al. (2002: open pentagons) with our data. Solid lines mark the one-to-one
relation. Table~11 summarizes the results of the comparison. } 
\label{fig6}
\end{figure*}

\section{Results} 

For each galaxy the set of 25 indices obtained for the 7 apertures and  the 4 gradients   are
provided in electronic form with the format shown in Tables 9 and 10 respectively. 
The structure of the above tables is the following: each aperture (or gradient) is described
by two rows. In the first raw: col.~1 gives the galaxy identification, col.~2 the number
of the aperture, col.~3 is a flag: 0 stands for values of indices,
col.~4 and Col.~5 give the radii delimited by the aperture, from col.~6 to 30 individual
indices are given. In the second row: col.~1 gives the galaxy identification, col.~2 the number
of the aperture, col.~3 is a flag: 1 stands for error of the indices,
col.~4 and Col.~5 give the luminosity weighted radius of the aperture and the
adopted equivalent radius, from col.~6 to 30 are given the errors of the indices.

In electronic form are also available, under request to the authors, the tables of raw indices 
(before the velocity dispersion correction) as well as the fully calibrated spectra 
(apertures and gradients) in digital from for each galaxy. 

In Figures 8-13 we show as examples the trend with radius of the Mg$_2$, Fe(5335) and
H$\beta$ indices for the 50 galaxies in the sample (apertures are marked with open
squares, gradient with dots).  

\subsection{Comparison with the literature} 

A significant fraction, about 60\%, of galaxies in the present sample has one previous 
measurement in the Lick--IDS system but basically  restricted within the r$_e$/8 region. 
Line-strength measures  obtained both from apertures and gradients outside this area
and within the r$_e$/8 region, with the present radial mapping, are completely new. 

The set of line--strength indices in the literature available for a comparison 
is quite heterogeneous since indices are measured within different apertures.
Furthermore, there are many possible sources of systematic errors from 
seeing effects to the calibration applied to shift indices to the same spectrophotometric 
system and to velocity dispersion correction.

Three galaxies, namely NGC 1453, NGC 4552 and  NGC 5846 are in the 
Gonz\' alez (\cite{G93}) sample. Three galaxies, namely NGC 1553, NGC 6958 
and NGC 7135 belong to the sample observed by Longhetti et al. (\cite{L98a}). 
Twenty one  galaxies are in the sample provided by  Trager et al. (\cite{Tra98}) namely 
NGC 128, NGC 777, NGC 1052, NGC 1209, NGC 1380, NGC 1407, NGC 1426, 
NGC 1453, NGC 1521, NGC 2749, NGC 2962, NGC 2974, NGC 3489, NGC 3607, 
NGC 3962, NGC 4552, NGC 4636, NGC 5077, NGC 5328, NGC 7332 and NGC 7377. 
Eleven galaxies are in the sample recently published  by Beuing et al. (\cite{Beu02}), 
namely IC 1459, IC 2006, NGC 1052, NGC 1209, NGC 1407, NGC 1553, NGC 5898, 
NGC 6868, NGC 6958, NGC 7007 and NGC 7192. 

A global comparison with the literature is shown shown is Figure~\ref{fig6}. In detail:
(1) with Longhetti et al. (\cite{L98a}) the comparison is made with indices computed
on the aperture of 2.5\arcsec radius; (2) with Gonz\' alez (\cite{G93}) on r$_e/8$
aperture and (3) with Trager et al. (\cite{Tra98})  with indices computed 
within standard apertures; (4) with Beuing et al. (\cite{Beu02}) with indices 
computed on the aperture with radius r$_{e}$/10, taking into account that these 
authors did not correct H$\beta$  for emission infilling.

In Table~11 we present a summary of the comparison with the literature. 
Both the offset and the dispersion for the various indices in the table are
comparable (or better) of those obtained on the same indices by 
Puzia et al.~(\cite{Puzia2002}) in their spectroscopic study of globular cluster.    

The comparison of our data with Trager et al. (\cite{Tra98}) for each index is in
general better than that with other authors and, in particular, with Beuing et
al. (\cite{Beu02}).  

The comparison with Trager et al. (\cite{Tra98}) shows a zero point shift for Mgb 
values, our data being larger although within the dispersion. A large shift is also 
shown by G4300 and Ca4227 both also visible in the comparison with Lick stellar
indices.  Beuing et al. (\cite{Beu02})  indices are on consistent range of
values, although some systematic effects and zero point offsets are present. While there
is a good agreement with the H$\beta$ (without emission correction), Mg$_1$, Mg$_2$ and 
Fe5335 line--strength indices, Mgb and  Fe5270 show a large zero point differences 
and dispersion. Beuing et al. \cite{Beu02} provided a comparison with 
Trager et al. (\cite{Tra98}), on a partially different sample. They show a basic 
agreement for the H$\beta$, Mg$_1$, Mg$_2$ and Mgb (although both a zero point shift and 
a different slope are visible, e.g. in H$\beta$ and Mg$_1$) while 
Fe5270, and Fe5335 indices show a quite large dispersion and  zero point
shift as shown in our Figure~\ref{fig6}.

\begin{figure}
\psfig{figure=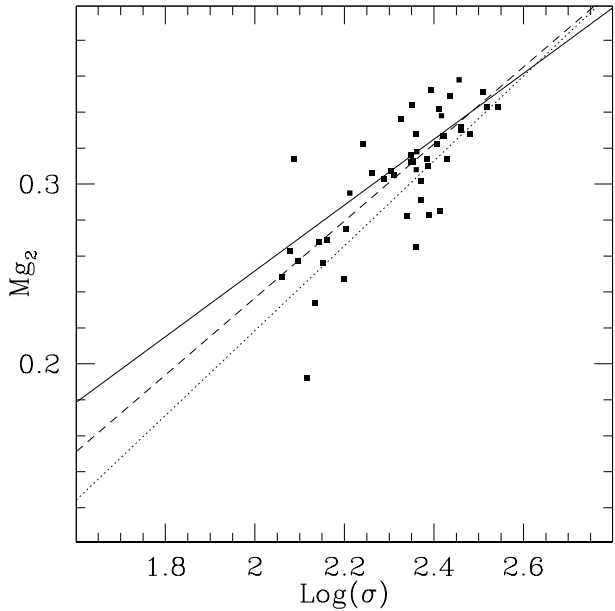,width=8.5cm,clip=}
\caption{Mg$_2$ versus $\sigma$ relation. Our Mg$_2$ line-strength indices measured
within the SLOAN aperture (r=1.5\arcsec) are plotted after the transformation 
to the Lick-IDS system. The solid dashed line is the least-square fit obtained 
Trager et al. ~(\cite{Tra98}). The  the dotted and long-dashed lines represent  
 the least-square fit obtained by Bernardi et al. (1998) for the field 
sample of 631 galaxies and  our fit (value at $\sigma_{300~km~s^{-1}}$ = 0.339,
slope of relation = 0.214) to the present data  respectively.}
\label{fig7}
\end{figure}

The well known Mg$_2$ vs. $\sigma$ relation is plotted in Figure \ref{fig7}. 
Our Mg$_2$ values computed at the r=1.5\arcsec, compatible with the SLOAN
apertures are plotted versus the corresponding velocity dispersion values. 
Adopting  the parameters of the fitting given in Worthey \& Collobert (\cite{WC03})
(their Table 1), the dotted line is the least-square fit obtained by Bernardi et al.
(\cite{Ber98}) on the sample of 631 field early-type galaxies while the
solid line is the Trager et al. (\cite{Tra98})  fit.
Bernardi et al. fit is made on the Mg$_2$ index computed using
the Lick definition but not transformed to the Lick-IDS system. The
long-dashed line shows our least-squares fits to the present data: notice that
the value of the Mg index at $\sigma$=300 km~s$^{-1}$ of 0.339  is well consistent 
with that of Trager et al. (\cite{Tra98}) while the slope for our small set of galaxies
is in between those given by the above considered authors.

\begin{figure*}
\psfig{figure=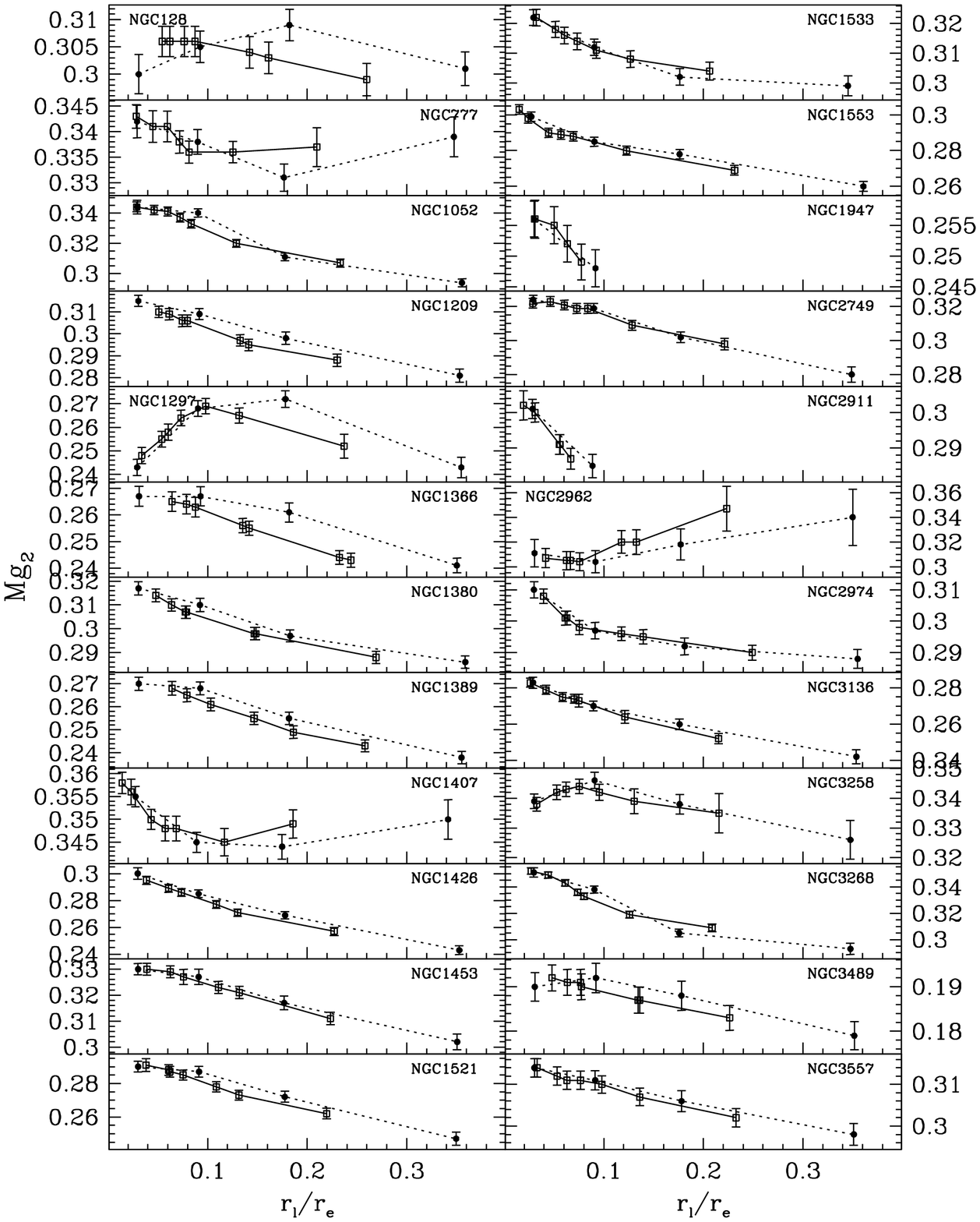,width=16cm,clip=}
\caption{Fully corrected Mg$_2$ line-strength index as function of the luminosity
weighted radius normalized to the galaxy equivalent radius R$_e$. Apertures are
indicated with open squares, while gradients are indicated with full dots.} 
\label{fig8}
\end{figure*}

\begin{figure*}
\psfig{figure=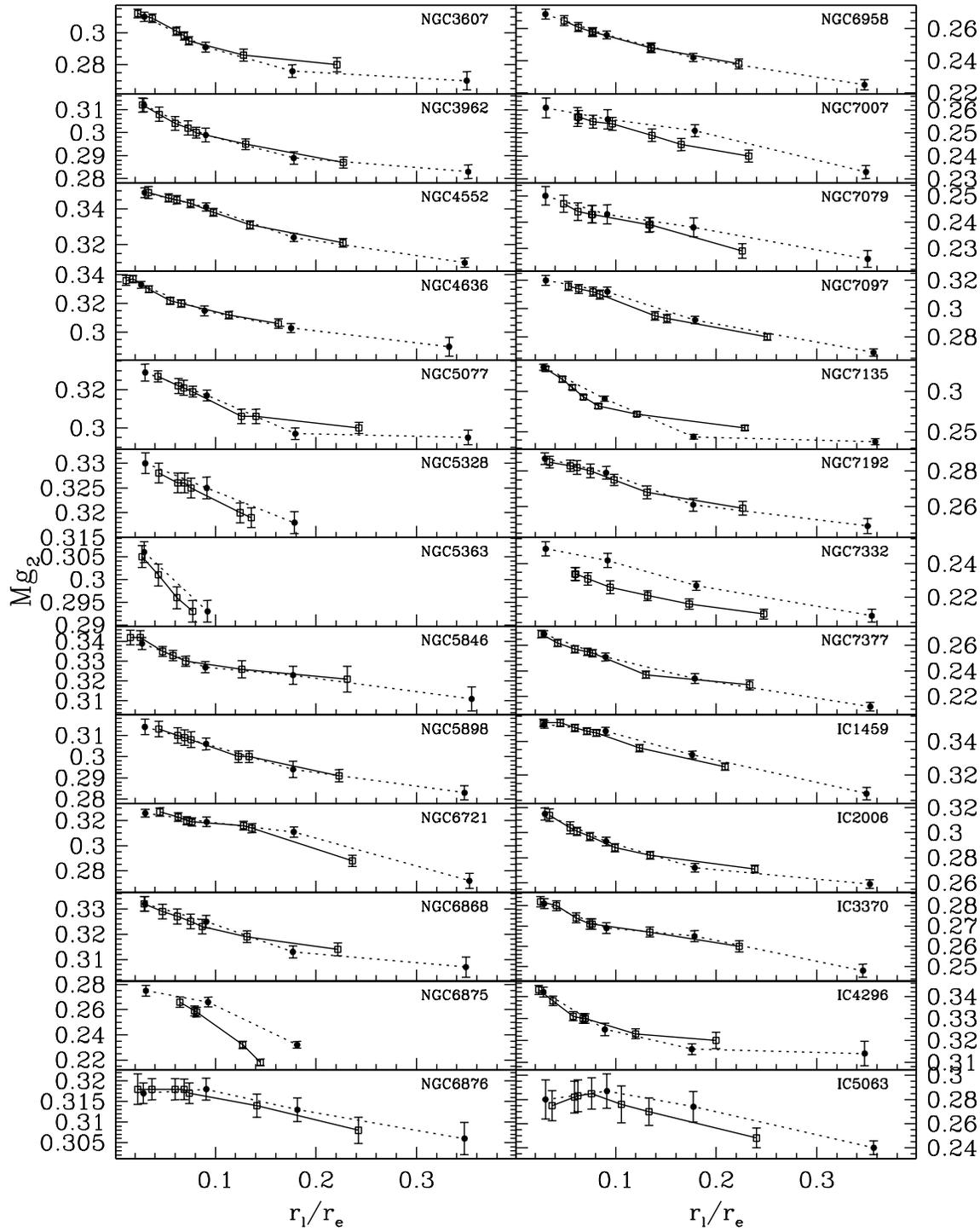,width=16cm,clip=}
\caption{As in Figure 8.} 
\label{fig9}
\end{figure*}
\begin{figure*}
\psfig{figure=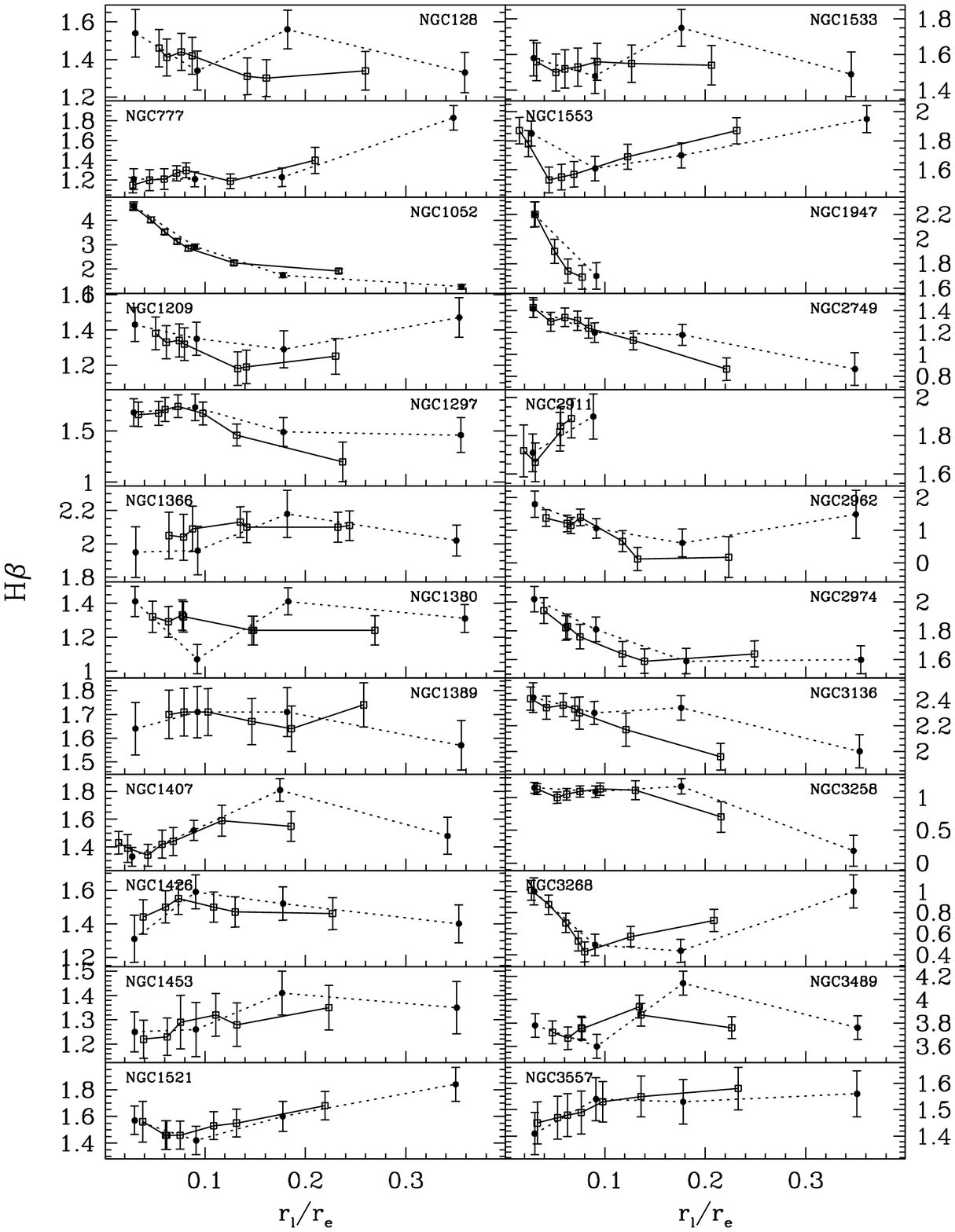,width=16cm,clip=}
\caption{Fully corrected H$\beta$ line-strength index as function of the luminosity
weighted radius normalized to the galaxy equivalent radius R$_e$. Apertures are
indicated with open squares, while gradients are indicated with full dots.} 
\label{fig10}
\end{figure*}
\begin{figure*}
\psfig{figure=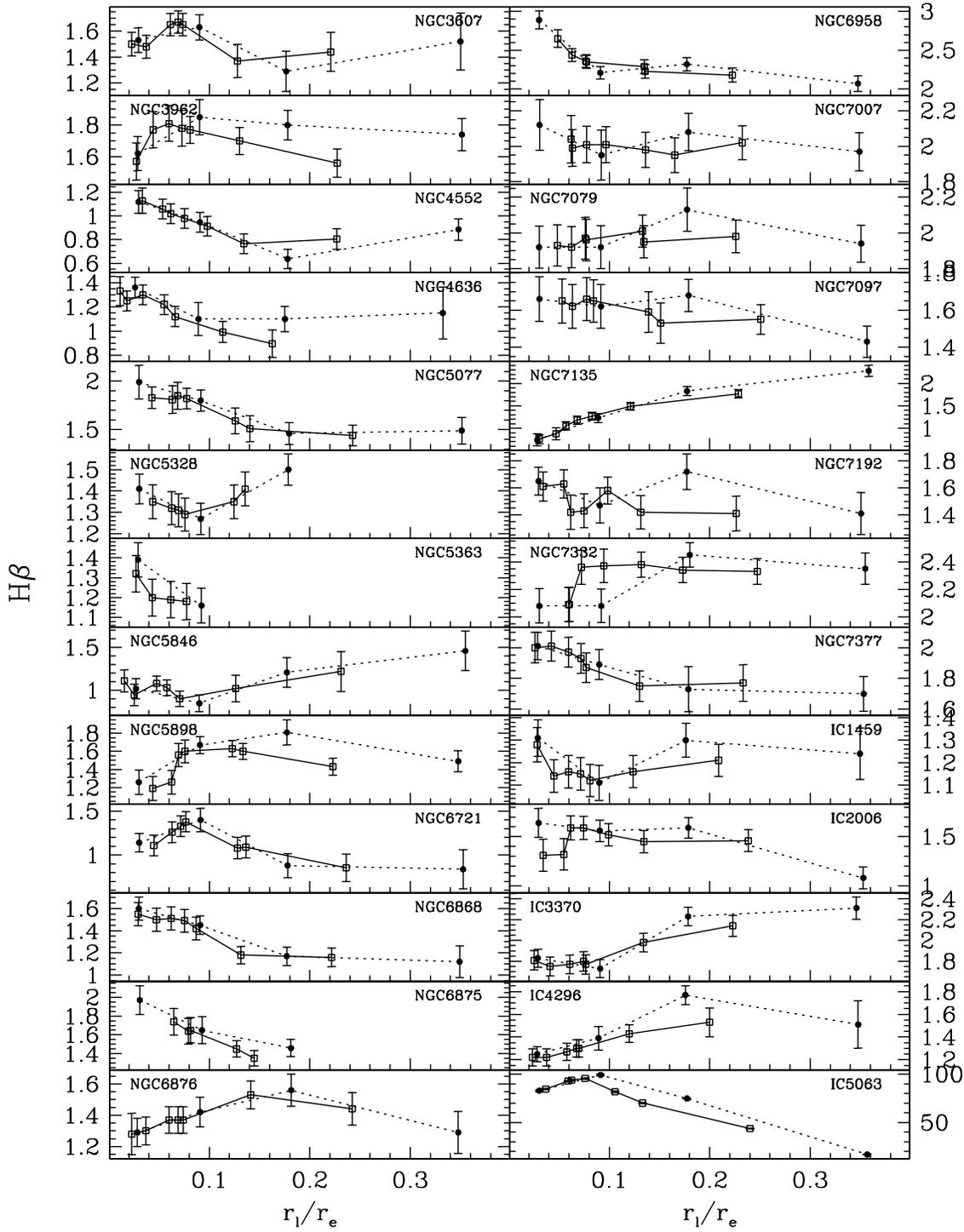,width=16cm,clip=}
\caption{As in Figure 10.} 
\label{fig11}
\end{figure*}
\begin{figure*}
\psfig{figure=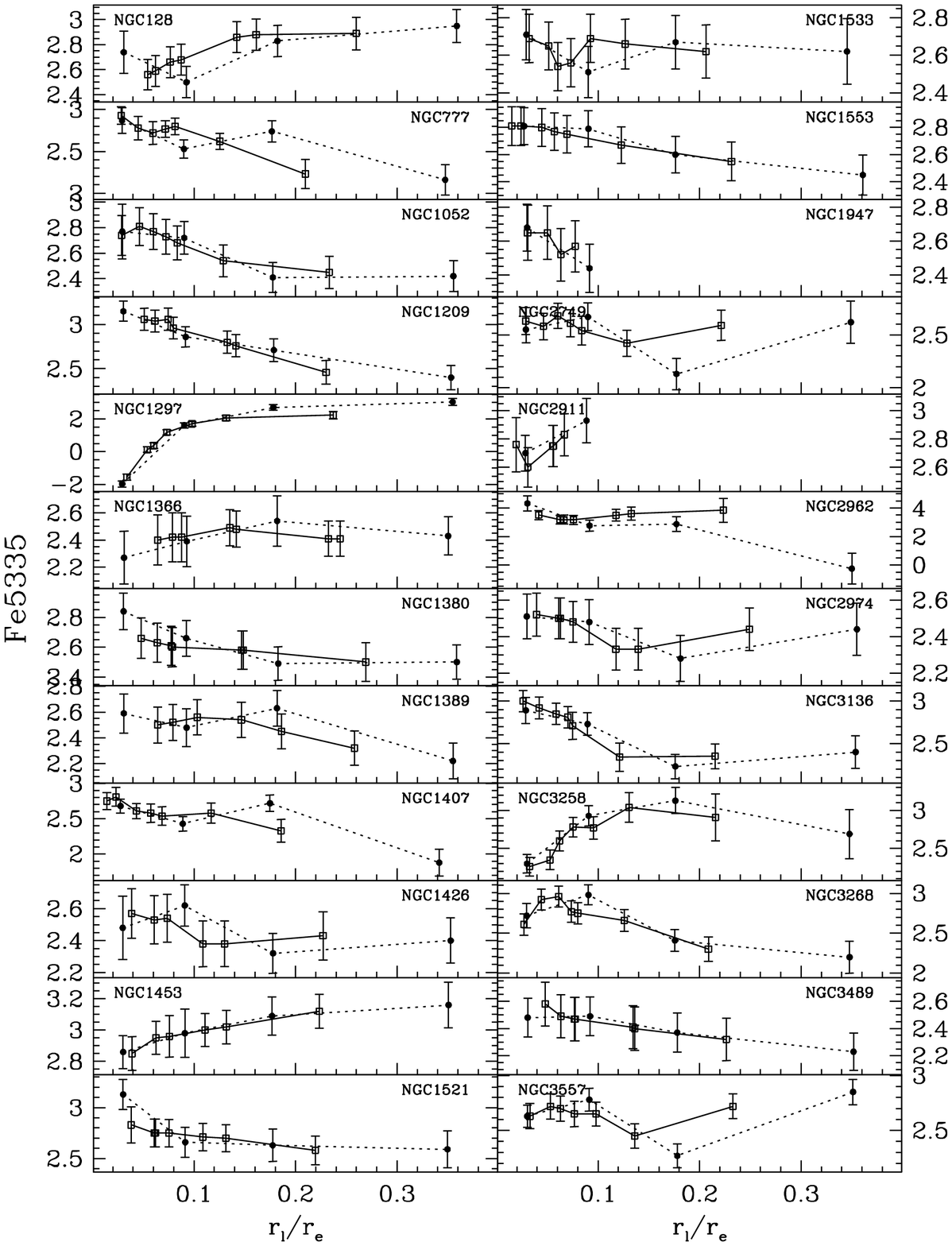,width=16cm,clip=}
\caption{Fully corrected Fe5335 line-strength index as function of the luminosity
weighted radius normalized to the galaxy equivalent radius R$_e$. Apertures are
indicated with open squares, while gradients are indicated with full dots.} 
\label{fig12}
\end{figure*}

\begin{figure*}
\psfig{figure=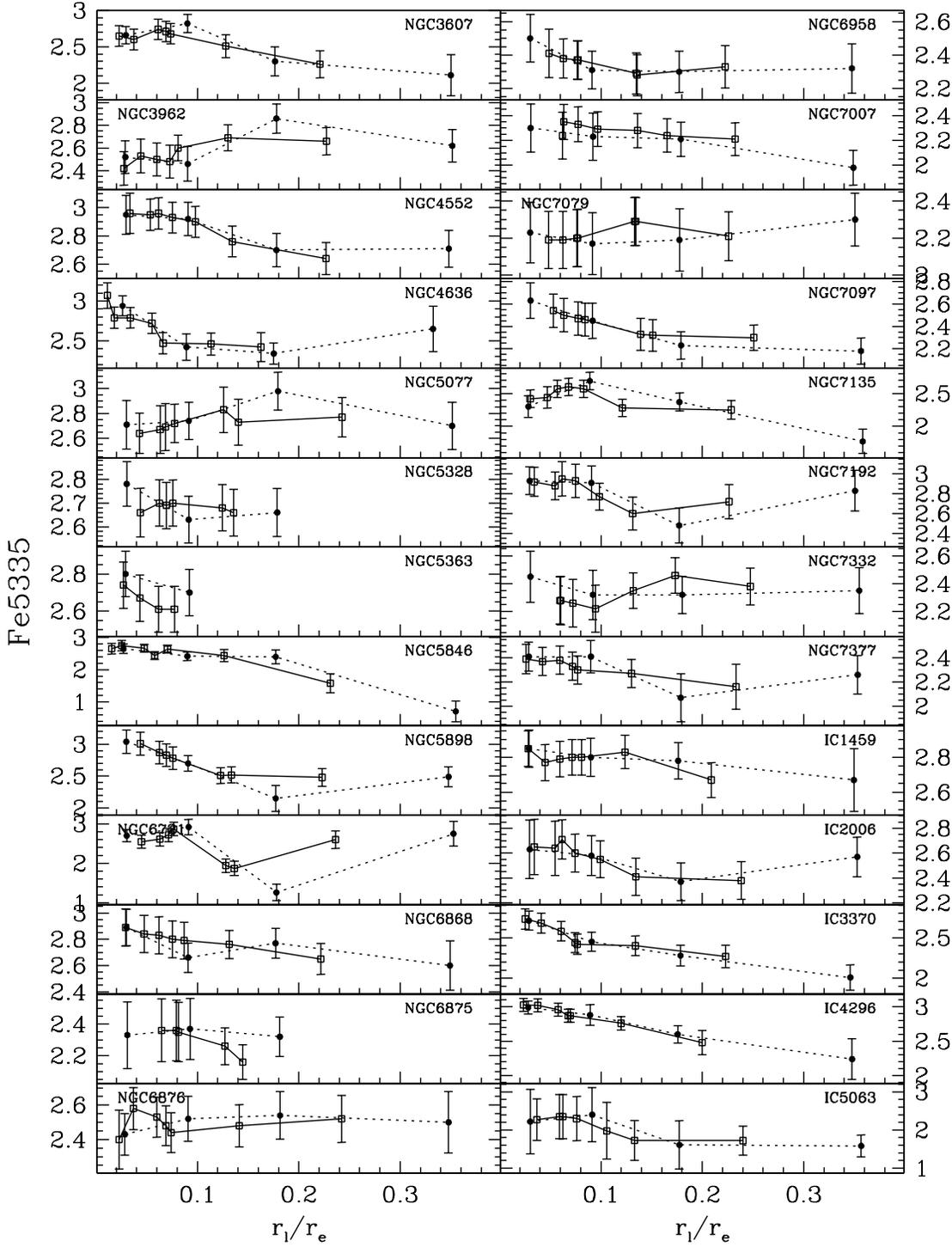,width=16cm,clip=}
\caption{As in Figure 12.} 
\label{fig13}
\end{figure*}

\begin{table}
\small{
\begin{tabular}{lcrrr}
& &  & &\\
\multicolumn{5}{c}{\bf Table 11. ~~~ Comparison with the literature} \\
\hline\hline
\multicolumn{1}{c}{index}
&\multicolumn{1}{c}{N$_{gal}$}
& \multicolumn{1}{c}{offset}
& \multicolumn{1}{c}{dispersion}
& \multicolumn{1}{c}{units}
 \\
 \hline
 & & G93 + Long98 & & \\
\hline
 & & & &\\
        Hb     &    6 &    0.171 &   0.886 & \AA\\
       Mg$_2$   &    6 &   -0.006 &   0.014  & mag \\
       Mgb   &    6 &    0.078 &   0.458 & \AA \\ 
    Fe5270  &    6 &    0.208 &   0.334 & \AA \\ 
    Fe5335  &    6 &   0.160 &   0.348 & \AA  \\ 
       Mg$_1$   &    6 &  -0.007 &   0.011 & mag \\ 
 \hline
 & & Trager et al. (1998) & & \\
\hline
        Hb    &   20   &  -0.031 &   0.405 & \AA \\
       Mg$_2$   &    20  &  0.004  &  0.016  & mag \\
       Mgb   &    20  &  0.218  &  0.321  &\AA \\
    Fe5270  &     21 &  0.051 &   0.333 & \AA \\
    Fe5335   &    21 &  -0.038  &  0.376 &\AA \\
       Mg$_1$    &   20  & -0.002   & 0.013 & mag \\
     G4300   &    21 &   0.463  &  0.946& \AA \\
    Ca4227   &    21 &   0.519  & 0.757 &\AA \\
 \hline
 & & Beuing et al. (2002)& & \\
\hline
        Hb    &   11 &   -0.059  &  0.271 &  \AA \\
       Mg$_2$   &    11 &   0.018  &  0.025 & mag \\       
       Mgb   &    11 &   0.470   & 0.628 &\AA \\
    Fe5270  &    11 &   0.339 &   0.603 &\AA \\
    Fe5335  &    11 &   0.075  &  0.455 &\AA \\
       Mg$_1$   &    11  &  0.011   & 0.018& mag \\
    \hline
\end{tabular}}
\label{table11}

\medskip
Offsets and dispersions of the residuals between our data and the literature. 
Dispersions are 1 $\sigma$ scatter of the residuals.
Beuing et al. (2002) do not compute G4300 and Ca4227 indices. For G4300 and Ca4227 indices
Gonz\' alez (1993) and Longhetti et al. (1998) used a (slightly) different definition than 
Trager et al. (1998)  and consequently comparisons are not reported in the table. The 
comparison for H$\beta$ index between our data and those of Trager et al. (1998) and
Beuing et al. (2002) are made using uncorrected data. 
\end{table}

\section{Summary}
This paper is the first in a series dedicated to the spectroscopic study of
early--type galaxies showing emission lines in their optical spectrum. It
presents the line--strength index measurements for 50 galaxies residing in
cosmological environments of different galaxy richness. The morphology and the
kinematics of the gas component with respect to the stellar population are
summarized using the available literature in order to characterize each galaxy
for subsequent studies. According to current views supported by numerical
simulations (see e.g. Colless et al. (\cite{Coll99})) a large fraction of the
galaxies in the sample display morphological and/or kinematical signatures of
past interaction/accretion events. For each galaxy, three integrated spectra for
different galactocentric distances, center, r$_e$/4 and r$_e$/2, have been
measured. Spectra, in the wavelength range of 3700\AA$< \lambda < $ 7250\AA,
have been collated in the Atlas which covers a large set of E and S0
morphological classes.

The paper is dedicated to the characterization of the emission galaxy
underlying stellar population through the preparation of a data base of their
line-strength indices in the Lick-IDS system once corrected for several
effects including infilling by emission and velocity dispersion. 

For each object we extracted 7 luminosity weighted apertures (with radii: 1.5\arcsec, 
2.5\arcsec, 10\arcsec,  r$_e$/10, r$_e$/8, r$_e$/4 and r$_e$/2) corrected for the galaxy 
ellipticity and  4 gradients (0 $\leq$ r $\leq$r$_e$/16, r$_{e}$/16 $\leq$ r $\leq$r$_e$/8,
r$_{e}$/8 $\leq$ r $\leq$r$_e$/4 and r$_{e}$/4 $\leq$ r $\leq$r$_e$/2). For
each aperture and gradient we measured 25 line--strength indices: 21 of the set
defined by the Lick-IDS ``standard'' system (Trager et al. \cite{Tra98}) and 4 
introduced by Worthey \& Ottaviani  (\cite{OW97}).

Line-strength indices, in particular those used to build the classic H$\beta$-$<
MgFe>$ plane, have been compared with literature. 
 A direct comparison was made with Gonz\' alez (\cite{G93}),
Longhetti et al. (\cite{L98a}), Beuing et al. (\cite{Beu02})
 but in particular with the larger data
sets of Trager et al. (\cite{Tra98})  showing
the reliability of our measures. 

In forthcoming papers we plan (1) to model  this information for investigating 
the ages and metallicities of the bulk of the stellar populations in these galaxies; 
(2) to extend the study calibrating and modeling other ''blue indices"  outside the 
Lick/IDS system  namely $\Delta$(4000\AA) (Hamilton
\cite{Ha85}) , H+K(CaII) and H$\delta$/FeI (Rose \cite{Ro84}; Rose
\cite{Ro85}) correlated with the age of the last starburst event in
each galaxy;  (3) to complete the study of the emission line component
 on a finer spatial grid.

\begin{acknowledgements} The authors would like to thank the referee,
Guy Worthey, whose comments and suggestions significantly contributed to
improve the paper. We are deeply indebted with Enrico V. Held for his
invaluable help.  We thank Daniela Bettoni, Nicola Caon and Francois
Simien for making their velocity dispersion measurements available to us
in digital form. RR acknowledges  the partial support of the ASI
(contract I/R/037/01). WWZ acknowledges the support of the Austrian
Science Fund (project P14783) and of the Bundesministerium f\"ur
Bildung, Wissenschaft und Kultur.
\end{acknowledgements}

\appendix
\section{Relevant notes on individual galaxies from the literature}
We report in this section some studies relevant to the present investigation
performed in the recent literature with particular attention to the properties 
of the ionized gas with respect to the bulk of the stellar component.

\underbar{NGC 128} Emsellem \& Arsenault (\cite{Em97}) present a
study of the gas (and dust) disk tilted at an angle of 26$^\circ$ with respect 
to the major axis of the galaxy.  The stellar and gas velocity fields show that
the angular momentum vectors of the stellar and gaseous
components are reversed, suggesting that the gas component orbits
suffer the presence of a tumbling bar, possibly triggered by the
interaction of NGC 128 with the nearby companion NGC 127.  The gas
extends at least up to 6\arcsec\ from galaxy center, and in the inner
parts line ratios are typical of LINER and consistent with the gas
being ionized by post-AGB stars.
D'Onofrio et al. (\cite{Do99}) evaluate the central gas mass is 
$\approx$ 2.7 10$^4$ M$_\odot$. The dust does not have 
the same distribution as the gas but is largely confined to the region
of interaction between NGC 128 and NGC 127. They calculated a 
dust mass of $\approx$ 6$\times$10$^6$ M$_\odot$ for NGC 128.

\underbar{NGC 777} The kinematics and the photometry of this galaxy 
were obtained by Jedrzejewski \& Schechter (\cite{JS89}). Both the P.A.
and the ellipticity profile appear nearly constant at about 149\degr\
and 0.18 respectively. Both the kinematics along the major and minor
axes have been investigated. A rotation of about 50 km~s$^{-1}$ is
measured along the major axis while no apparent rotation is detected
along the minor one. 

\underbar{NGC 1052} The galaxy is known as a prototypical LINER
(Heckman \cite{He80}). Plana et al. (\cite{Pl98}), performing
Fabry-Perot observations, succeeded in disentangling two gas
components both kinematically decoupled from the stellar
component. Both, in fact, have their apparent major axis nearly
perpendicular to the stellar one. The ionized gas of the main
component was detected up to 30\arcsec\ from the center, while the
second one extends up to 15\arcsec. The main component shows rotation
with an apparent major axis of 45$^\circ \pm$ 4$^\circ$ similar to
that of the HI emission detected by van Gorkom et al. (\cite{vG86})
with which it shares also similar kinematical characteristics. The
velocity field of both components presents shapes and velocity
dispersion in agreement with models of inner disks found in elliptical
galaxies.
Recently Gabel et al.  (\cite{Ga00}) imaged the central part of the
galaxy using WFPC2/HST with an H$\alpha$ filter showing that a
filamentary nebular emission extends about 1\arcsec\ around a compact
nucleus with a more diffuse halo extending to further distances. At
the position angle of $\approx$ 235$^\circ$ there is a narrow filament
of H$\alpha$ emission. A radio jet/lobe at a position angle of
$\approx$ 275$^\circ$ has been evidenced by Wrobel \& Heeschen
(\cite{Wro84}). The emission line region is much more extended, as
discussed above. Gabel et al. (\cite{Ga00}) examined whether or not
the ionizing continuum flux is sufficient to power the above extended
emission line region. They conclude that a pure-central source
photo-ionization model with the simplest non-thermal continuum (a
simple power law) reproduces the emission line flux in the inner
region of NGC 1052. Other processes, such as shocks or photoionization
by stars, are not required to produce the observed emission. However,
the contribution of these latter mechanisms cannot be ruled out in the
extended nebular emission region.
Recently Raiman et al. (\cite{Ra01}) have analyzed the spectra of NGC
1052 and IC 1459, classified as LINERS, at several galactocentric
distances from the nucleus.  They found that these objects have both
the nucleus redder than the surroundings and nuclear absorption lines
stronger than outside the nucleus similarly to normal galaxies. On the
other side the spectral synthesis of NGC 1052 and IC 1459 indicate
that they have only a $\approx$ 10 - 20 \% larger contribution of the
1-Gyr component at the nucleus with respect to normal galaxies which
are dominated by the old metal rich component whose contribution is
decreasing outwards. The above authors exclude the presence of young
massive stars found by Maoz et al. (\cite{Maoz98}).

\underbar{NGC 1209} The galaxy is part of the group dominated by NGC
1199.  The surface photometry and the geometrical study, performed by
Capaccioli et al. (\cite{CPR88}), extends up to $\mu_B \approx$ 28
mag~arcsec$^{-2}$. The ellipticity grows from 0.22 to 0.57 with
basically no twisting ($<P.A>$ = 51$^\circ$ $\pm$2$^\circ$ up to
126\arcsec) suggesting that the galaxy is an S0.

\underbar{NGC 1380} D'Onofrio et al. (\cite{Do95}) suggest the presence
of a dusty nucleus. The stellar component has a strong gradient in both
rotation and velocity dispersion curves and the disk dominates
outside 20\arcsec. Kuntschner (\cite{Ku00}), studying stellar populations of
early-type galaxies in the Fornax cluster, found that this bright
lenticular has properties similar to those of ellipticals suggesting
that they experienced similar star formation histories. NGC~1380
exhibits an overabundance in magnesium compared to iron, similarly to
most ellipticals in the Fornax cluster. 

\underbar{NGC 1389} This galaxy belongs to the Fornax
cluster. Phillips et al. (\cite{Phi96}) examined the nucleus of the
galaxy using the HST Planetary Camera, finding no evidence for an
unresolved central point source. The image shows a smooth light
distribution sharply peaked at the center and isophote twisting
within 5\arcsec.

\underbar{NGC 1407} The (B-V) image of this galaxy, which is 
a radio source, reveals a circumnuclear ring slightly redder than 
the nucleus (Goudfrooij \cite{Gou94}).
The stellar kinematics were studied along several axes by Longo et al. (\cite{Lon94}).
Franx et al. (\cite{Fra89}) found  significant rotation along the minor axis.

\underbar{NGC 1426} The study performed by Capaccioli et al. (\cite{CPR88}) 
extends up to $\mu_B \approx$ 28 mag~arcsec$^{-2}$ and indicates that 
the galaxy is an S0 with basically no isophotal twisting 
($<P.A>$ = 105$^\circ$ $\pm$1$^\circ$ up to
100\arcsec).  Quillen (\cite{Qui00}) have observed the galaxy with HST
NICMOS. The central 10\arcsec\ show very regular isophotes with no
twisting or deviations from elliptical shapes. The core has a power
law profile and no dust features are detected. The extended
rotation curve and velocity dispersion profiles, obtained by Simien \&
Prugniel (\cite{SP97a}), do not show any peculiarities.

\underbar{NGC 1453} Pizzella et al. (\cite{Pi97}) found that this E2 galaxy has
a twisting of 10\degr. The  H$\alpha$ image reveals the presence of an
ionized disk misaligned with respect to the stellar isophotes by
$\approx$ 56\degr\ suggesting an intrinsic triaxial shape.

\underbar{NGC 1521} The surface photometry and the geometrical study
performed by Capaccioli et al. ( \cite{CPR88}) extends up to $\mu_B \approx$ 27-28
mag~arcsec$^{-2}$ at 3.8 r$_e$. The galaxy shows a peculiar light distribution
with a change in slope along the major axis and a significant twisting
($\approx$=20$^\circ$  up to 126\arcsec). The extended rotation curve
and velocity dispersion profiles, obtained by Simien \& Prugniel (\cite{SP97a}),
do not show peculiar features.

\underbar{NGC 1553} The galaxy belongs to the shell galaxies sample of of Malin
\& Carter (\cite{MC83}). The galaxy kinematics is typical of an early S0s and
shows two maxima in the rotations curve (Kormendy \cite{Kor84}; Rampazzo
\cite{R88}; Longhetti et al. \cite{L98b}). No rotation is found along the minor
axis. The H$\alpha$ narrow band image of this galaxy (Trinchieri et al.
\cite{Tri97}) shows a strong nuclear peak and a bar-like feature $\approx$ in
the North-South direction that ends in  spiral structure at 8\arcsec\ from the
nucleus. Blanton et al.  (\cite{Bla01}) proposed using {\it Chandra} data that
the center of NGC~1553 is probably an obscured AGN while,  the X-ray diffuse
emission exhibits significant substructure with a spiral feature passing through
the center of the galaxy.
Longhetti et al. (\cite{L98a,L99,L00}) studied the stellar population of this galaxy
suggesting that the age of a secondary burst  is old, probably associated to the
shell formation. Rampazzo et al. (\cite{R03}) measured the velocity field
of the gas component using Fabry-Perot data. In the central region 
of NGC 1553 the ionized gas is co-rotating with the stellar component.

\underbar{NGC 1947} The galaxy is considered a minor-axis dust-lane elliptical
(Bertola et al. \cite{BBZ92b}). The stellar component of this galaxy is rotating
around the minor axis, perpendicularly to the gas rotation axis. The gas
component forms a warped disk whose external origin is suggested by Bertola et
al. (\cite{BBZ92b}).

\underbar{NGC 2749} The galaxy, studied by Jedrzejewski \& Schechter
(\cite{JS89}), shows strong rotation ($\geq$ 100 km~s$^{-1}$) along
both the major and minor axes. A measure of the gas rotation
curve was attempted using the $\lambda$ 5007 \AA\ line. They suggested that the gas
is not rotating, with an upper limit of the order of one-half the stellar rotation velocity on either axis.

\underbar{NGC 2911} Known also as Arp 232 the galaxy is classified as
a LINER in the V\`eron-Cetty \& V\`eron (\cite{Ver01}) catalog. Michard \&
Marchal (\cite{Mich94}) suggest that this is a disk dominated S0 with
a significant dust component.

\underbar{NGC 2962} The extended rotation curve
and velocity dispersion profiles, obtained by  Prugniel \& Simien (\cite{PS00}),
do not show peculiarities.

\underbar{NGC 2974} This E4 galaxy imaged in H$\alpha$ reveals the
presence of an ionized disk misaligned with respect to the stellar
isophotes by $\approx$ 20\degr\ (Pizzella et al. \cite{Pi97};
Ulrich-Demoulin et al. \cite{U84}; Goudfrooij \cite{Gou94}). The
galaxy has an HI disk (Kim \cite{K89}) with the same
rotation axis and velocity as the inner ionized one. Plana et
al. (\cite{Pl98}) suggest that this object is a good candidate for an
internal origin of the ionized gas.  Bregman et al. (\cite{Bre92})
present evidence of a spiral arm structure and Cinzano \& van der
Marel (\cite{CvM94}) could not discard the hypothesis that NGC 2974 
is a Sa galaxy with an unusually low disk-to-bulge ratio.

\underbar{NGC 3136} Using an H$\alpha$+[NII] image Goudfrooij (\cite{Gou94})
detected  an extended emission with a maximum at the nucleus and peculiar
arm-like feature extending out to $\approx$ 55\arcsec\ from the center. Dust
absorption is  found to be associated with the ionized gas. Koprolin \&
Zeilinger (\cite{KZ00}) suggest that a counterrotating disk with a dimension of
2\arcsec\ is located  4\arcsec\  from the galaxy center.

\underbar{NGC 3258} Koprolin \& Zeilinger \cite{KZ00} measured
a very low rotation velocity of 39$\pm$10 km~s$^{-1}$ for this galaxy.

\underbar{NGC 3268} Koprolin \& Zeilinger (\cite{KZ00}) found that the galaxy
has an asymmetric rotation curve with respect to the nucleus, probably due to
the presence of a dust-lane.

\underbar{NGC 3489} Gas and stars show a fast rotation along the major axis of
the gas distribution which roughly coincides with the major axis of the stellar
isophotes. The gas shows  rotation along the minor axis while no stellar
rotation is measured. There is evidence for a distinct nuclear stellar component
(within r $\approx$ 3\arcsec) (Caon et al. \cite{CMP00}).

\underbar{NGC 3557} The color map reveals a possible ring of dust near the
center of the galaxy (Colbert et al. \cite{C01}). The galaxy is a double
tail radio source with a central knot and a jet (Birkinshaw \& Davies \cite{Bi85}).
Goudfrooij (\cite{Gou94}), using an H$\alpha$+[NII] image, shows that
the outer isophotes of the line emission twist gradually toward the apparent
major axis of the galaxy.

\underbar{NGC 3607} Caon et al. (\cite{CMP00}) observe stellar kinematics along
the major axis which is also the major axis  of the gas distribution.
The gas rotation curve has a steeper gradient and a larger amplitude
than the stellar one.

\underbar{NGC 3962} Birkinshaw \& Davies (\cite{Bi85}) revealed a radio source
in the center of the galaxy.  The morphology and the kinematics of the
ionized gas confirm the presence of two distinct subsystems: an inner
gaseous disk and an arc-like structure. The inner gaseous disk shows
regular kinematics with a major axis near P.A.= 70\degr\ and inclination
of about 45\degr\ (Zeilinger et al. \cite{Z96}).

\underbar{NGC 4552} The extended kinematics of this galaxy has been recently
obtained by Simien \& Prugniel (\cite{SP97b}) who 
measure a very low maximum rotation velocity of 17$\pm10$ km~s$^{-1}$. 
NGC 4552 is a member of the Malin \& Carter (\cite{MC83}) supplementary list of
galaxies showing shells (they report "two or three shells and jet")

\underbar{NGC 4636} Caon et al. (\cite{CMP00}) observed the galaxies along three axes
and found that the gas has very irregular velocity curves. Zeilinger et al. (\cite{Z96})
suggested that the gas could suffer for turbulent motions due to material not yet
settled.

\underbar{NGC 5077} Caon et al. (\cite{CMP00}) found that the galaxy exhibits a
gaseous disk  with major axis roughly orthogonal to the galaxy
photometric major axis. The gas isophotes show a twisting and
a warp (Pizzella et al. \cite{Pi97}). The gas has a symmetric
rotation curve with an amplitude of 270 km~s$^{-1}$ at r=13\arcsec.
Along this axis the stellar rotation is slow. Along the axis at
P.A.=10$^\circ$, the stellar velocity curve shows a counterrotation in
the core region (r$<$5\arcsec) with a corresponding peak in
the stellar velocity dispersion.

\underbar{NGC 5266} This galaxy has a dust lane along the apparent minor
axis of the elliptical stellar body. The kinematics of NGC 5266 has
been extensively studied by Varnas et al. (\cite{Va87}) revealing a
cylindrical rotation of the stellar component (V$_{max}$=212 $\pm$ 7
km~s$^{-1}$) about the short axis and smaller rotation (V=43 $\pm$ 16
km~s$^{-1}$) about the long axis. The stellar velocity rotation is
210 $\pm$ 6 km~s$^{-1}$ and decreases with radius to 
100 km~s$^{-1}$ at r $\approx$ 20\arcsec. The gas associated with the dust
rotates about the major axis of the galaxy with a velocity of 260
$\pm$ 10 km~s$^{-1}$.  In the warp the gas motion are prograde with
respect to the major axis stellar rotation. HI radio observations
reveal the presence of a large amount of cold gas probably distributed
in a rotating disk.

\underbar{NGC 5363} The galaxy has a  warped dust lane confined to the central
part along its apparent minor axis. Differently from NGC 5128,  the gas motions
in the warp are found to be retrograde with respect to the stellar body. 
Bertola et al. (\cite{Ber85}) suggest that is an indication that the warp is a
transient feature and of the external origin of the gas and dust system.

\underbar{NGC 5846} Ulrich-Demoulin et al. (\cite{U84}) have studied the ionized gas
component in this galaxy. Several studies suggests the presence of dust 
(see Goudfrooij \cite{GT98}) in the galaxy. 
Caon et al. (\cite{CMP00}) found that the gas shows
 an irregular velocity profile, while stars have very slow rotation.
 
\underbar{NGC 5898} Caon et al. (\cite{CMP00}) analyzed the stellar and the gas
kinematics up to 45\arcsec\ showing the existence of a stellar core of 5\arcsec\
in radius, aligned with the major axis, which counterrotates with respect to the
outer stellar body.  The ionized gas counterrotates with respect to the inner
stellar core and co-rotates with respect to outer stellar body. At the same
time, the gas counterrotates along the minor axis, indicating that the angular
momentum vectors of the stars and of the gas are misaligned, but not
anti-parallel. A moderate quantity of dust is also present.

\underbar{NGC 6721} Bertin et al. (\cite{B94}) obtained extended stellar
kinematics (to 0.8 r$_e$) for this object finding a sizeable rotational
velocity $\approx$ 120 km~s$^{-1}$. 

\underbar{NGC 6868} The Fabry-Perot study of Plana et
al. (\cite{Pl98}) shows that the line--of--sight velocity field of the
ionized gas component has a velocity amplitude of $\pm$ 150
km~s$^{-1}$. Caon et al. (\cite{CMP00}) show that along the axes at
P.A.=30$^\circ$ and 70$^\circ$ the gas and stars have similar
kinematical properties, but along P.A.=120$^\circ$ the gas
counterrotates with respect to the stellar component. Zeilinger et
al. (\cite{Z96}) noticed the presence of an additional inner gas
component which suggested could be due to the superposition of two
unresolved counterrotating components, one
dominating the inner region, the other dominating the outer
parts. Also stars show a kinematically--decoupled counterrotating
core. The stellar velocity dispersion decreases towards the galaxy center.

\underbar{NGC 6958} This galaxy belongs to the list of Malin \& Carter (\cite{MC83})
of southern shell early-type. Saraiva et al. (\cite{SFP99}) detected isophotal
twisting of about 100\degr\ ($\approx$ 70\degr\ in the
inner 5\arcsec) but no particular signature of interaction in the isophote
shape which is elliptical. They conclude that if the galaxy suffered
interaction, the companion galaxy has probably already merged.

\underbar{NGC 7007} Pizzella et al. (\cite{Pi97}) found that the disk is
misaligned by about 30\degr\ with respect to the stellar isophotes with
an inclination of 57\degr. The ionized gas disk counterrotates with
respect to the stellar body and a bow-shape dust lane is also visible
on the eastern side (Bettoni et al. \cite{Be01}).

\underbar{NGC 7079} A counterrotating disk-like structure of ionized
gas within 20\arcsec\ from the center has been detected by Bettoni \&
Galletta (\cite{BG97}). The stellar body kinematics is typical of an
undisturbed disk. Cool gas (CO) has been detected by Bettoni et
al. (\cite{Be01}).  The cool gas component shares the same kinematics
of the ionized gas.

\underbar{NGC 7097} Caldwell et al. (\cite{Ca86}) found that the gas
and stellar components counterrotate. Zeilinger et al. (\cite{Z96}) show
that the rotation curve of the gaseous disk have a steep central
gradient with a discontinuity in the central part which may be related
to the counterrotating stellar component.

\underbar{NGC 7135} The galaxy belongs to the list of shell galaxies
in the southern hemisphere compiled by Malin \& Carter
(\cite{MC83}). Longhetti et al. (\cite{L98a,L98b,L99,L00}) studied the
inner kinematics and the stellar population of this galaxy. Rampazzo
et al. (\cite{R03}) show that the gas corotates with the stellar body.

\underbar{NGC 7192} Carollo \& Danziger (\cite{CD94}) showed that the innermost 
8\arcsec\ region counterrotates with respect to the galaxy body at greater
radii. The authors  report that in correspondence with the kinematically
decoupled core an enhancement in the Mg$_2$ index is observed, while iron lines
are only weakly enhanced with respect to measurements at greater radii. The
surface photometry shows that the galaxy has a very regular,  round structure.

\underbar{NGC 7332} Plana \& Boulesteix 
(\cite{PB96}), using a Fabry-Perot (CIGALE), found two gas components
(see also NGC 1052). The velocity field is consistent with two counterrotating
emission systems.

\underbar {IC 1459}
This giant elliptical has a massive counterrotating stellar core
(M$\approx$10$^{10}$ M$_\odot$; Franx \& Illingworth (\cite{FIl88})
which hosts a compact radio source. The galaxy is also crossed by a
disk of ionized gas (see \S 2.1), whose emission is detected out to
35\arcsec\ and rotates in the same direction as the outer
stellar component but at a higher speed (350 km~s$^{-1}$). Therefore
counterrotation in this galaxy seems confined to the inner core and
affects only stars. Bettoni et al. (\cite{Be01}) detect
$^{12}$CO(J=2-1) emission decoupled both from the ionized gas and the
counterrotating stellar core, since the velocity centroid of the CO
emission is redshifted by about 100 km~s$^{-1}$ with respect to the
galaxy systemic velocity. Raimann et al. (\cite{Ra01}) presented
spectral syntheses at several galacto-centric distances (see notes for NGC 1052).

\underbar{IC 2006} A faint emission of ionized gas, characterized by a
velocity gradient which is smaller and inverted with respect to that
of stars, is detected within 25\arcsec\ of the nucleus. The
counterrotating ionized gas disk is highly turbulent with a measured
velocity dispersion of 190 km~s$^{-1}$ (Schweizer et
al. \cite{Sch89}). Kuntschner (\cite{Ku00}) found that IC 2006 has
stronger metal-line absorption than what would be expected from the
mean index-$\sigma_0$ relation for ellipticals, although from their
data it is not clear if the galaxy is too metal-rich or whether the
central velocity dispersion is lower than for other ellipticals
of the same mass.

\underbar{IC 3370} The galaxy is a box shaped elliptical, with a
prominent dust-lane in the inner region, showing evidence of
cylindrical rotation and X-shaped isophotes. Van Driel et
al. (\cite{VD00}) postulated that it is a candidate polar-ring
galaxy.

\underbar{IC 4296} The galaxy kinematics have been studied by Franx et
al. (\cite{Fra89}) and more recently by Saglia et al. (\cite{S93}) out
to 0.8r$_e$ confirming the counterrotating core, detected by previous authors.  
Saglia et al. suggest the presence of a diffuse dark matter halo.

\underbar{IC 5063}
Long--slit spectroscopy for the radio galaxy IC 5063 has uncovered a clear rotation
pattern for the gas close to the nucleus and a flat rotation curve further out,
to at least 19 kpc (Bergeron et al. \cite{Be83}). The velocity difference between the
two flat parts of the rotation curve on both sides of the nucleus is
$\Delta$V = 475 km~s$^{-1}$.
Colina et al. (\cite{Col91}) report a very high excitation emission
line spectrum for this early-type galaxy which hosts a Seyfert 2
nucleus that emits particularly strongly at radio wavelength. The high
excitation lines are detected within 1-1.5\arcsec\ of both sides of
the nucleus, which is approximately the distance between the radio core and 
both of the lobes. These lines indicate the presence of a powerful and hard
ionizing continuum in the general area of the nucleus and the radio
knots in IC 5063. It has been estimated (Morganti et al. \cite{Mo98})
that the energy flux in the radio plasma is an order of magnitude
smaller than the energy flux emitted in the optical emission
lines. The shocks associated with the jet-ISM interaction are,
therefore, unlikely to account for the overall ionization, and the NLR
must be, at least partly, photo-ionized by the nucleus, unless the
lobe plasma contains a significant thermal component.


\begin{thebibliography}{}
\bibitem[1996]{Ba96} Barnes, J. 1996 in {\it Galaxies: Interactions and Induced Star Formation}, Saas-Fee Advanced Course 26. Lecture Notes 1996. Swiss Society for Astrophysics and Astronomy, XIV, Springer-Verlag Berlin/Heidelberg, p. 275
\bibitem[1983]{Be83} Bergeron, J., Durret, F., Boksenberg, A. 1983, A\&A, 127, 322
\bibitem[1998]{Ber98}Bernardi, M., Renzini, A., Da Costa, L.N. et al. 1998, ApJ, 508, L43
\bibitem[1994 ]{B94} Bertin, G., Bertola, F., Buson, L. et al. 1994, A\&A, 292, 381
\bibitem[1985]{Ber85} Bertola, F., Galletta, G., Zeilinger, W.W. 1985, ApJ,  292, L51
\bibitem[1992a]{BBZ92a} Bertola, F., Buson, L.M.,  Zeilinger, W.W. 1992, ApJ, 401, L79
\bibitem[1992b]{BBZ92b} Bertola, F., Galletta, G., Zeilinger, W.W. 1992, A\&A, 254, 89
\bibitem[1997]{BG97} Bettoni, D., Galletta, G. 1997, A\&AS, 124, 61
\bibitem[2001]{Be01} Bettoni, D., Galletta, G. Garcia-Burillo, S., Rodriguez-Franco, A. 2001, A\&A, 374, 421
\bibitem[2002]{Beu02} Beuing, J., Bender, R., Mendes de Oliveira, C.,
Thomas, D., Maraston, C. 2002, A\&A, 395, 431.
\bibitem[1994]{Bin94} Binette, L., Magris, C.G., Stasinska, G., Bruzual, A.G. 1994, A\&A, 292, 13
\bibitem[1985]{Bi85} Birkinshaw, M., Davies, R.L.  1985, ApJ, 291, 32
\bibitem[2001]{Bla01}  Blanton, E. L.; Sarazin, C. L.; Irwin, J. A. 2001, ApJ, 552, 106
\bibitem[1992]{Bre92} Bregman, J.N., Hogg, D.E., Roberts, M.S. 1992, ApJ, 397, 484.
\bibitem[1993]{Bus93} Buson, L.M., Sadler, E.M., Zeilinger, W.W.  et al., 1993, A\&A, 280, 409
\bibitem[1992]{Buz92} Buzzoni, A., Gariboldi, G., Mantegazza, L. 1992, AJ, 103, 1814
\bibitem[1994]{Buz94} Buzzoni, A., Mantegazza, L. Gariboldi, G. 1994, AJ, 107, 513
\bibitem[1986]{Ca86} Caldwell, N., Kirshner R.P., Richstone, D.O. 1986, ApJ, 305, 136
\bibitem[2000]{CMP00} Caon N., Macchetto, D., Pastoriza, M. 2000, ApJS, 127, 39 
\bibitem[1988]{CPR88} Capaccioli,M., Piotto, G., Rampazzo, R. 1988, AJ, 96, 487  
\bibitem[1993]{CDB93} Carollo, M., Danziger, I.J., Buson, L.M. 1993, MNRAS, 265, 553 
\bibitem[1994]{CD94} Carollo, M., Danziger, I.J. 1994, MNRAS, 270, 523
\bibitem[1995]{Car95} Carrasco, L., Buzzoni, A., Salsa, M., Recillas-Cruz, E. 1995, in {\it Fresh Views on Elliptical Galaxies}, ed.s A. Buzzoni, A. Renzini, A. Serrano, ASP Conf. Series 86, 175 
\bibitem[1994]{CvM94} Cinzano, P., van der Marel, R.P. 1994, MNRAS, 270, 325
\bibitem[2001]{C01} Colbert, J.W., Mulchaey, J.S., Zabludoff, A.I.  2001, AJ, 121, 808
\bibitem[1991]{Col91} Colina, L., Sparks, W.B., Macchetto, F. 1991, ApJ, 370, 102
\bibitem[1999]{Coll99} Colles, M., Burstein, D.,  Davies, R., McMahan, R.K.,  
saglia, R.P., Wegner, G. 1999, MNRAS, 303, 813
\bibitem[1991]{RC3} De Vaucouleurs, G., de Vaucouleurs, A., Corwin, H.G. Jr. et al. 1991 {\it Third Reference Catalogue of Bright Galaxies}, Springer-Verlag, New York
\bibitem[1987]{DD87} Djorgovsky, S., Davis, M. 1987, ApJ, 313, 59
\bibitem[1995]{Do95} D'Onofrio, M., Zaggia S.R., Longo, G., Caon, N., Capaccioli, M. 1995, A\&A, 296, 319 
\bibitem[1999]{Do99} D'Onofrio, M., Capaccioli, M., Merluzzi, P., Zaggia, S., Boulesteix, J. 1999, A\&AS, 134, 437
\bibitem[1997]{Em97} Emsellem, E., Arsenault, R. 1997, A\&A, 318, L39
\bibitem[1989]{Fra89} Franx, M., Illingworth, G.D., Heckman, T. 1989, ApJ, 344, 613
\bibitem[1988]{FIl88} Franx, M., Illingworth, G.D. 1988, ApJ, 327, L55
\bibitem[1994]{FI94} Fried, J.W., Illingworth, G.D. 1994, AJ, 107, 992
\bibitem[2000]{Ga00} Gabel, J.R., Bruhweiler, F.C., Crenshaw, D,M., Kraemer, S.B., Miskey, C.L. 2000, ApJ, 532, 883
\bibitem[1993]{G93} Gonz\' alez, J.J. 1993, Ph.D. thesis , Univ. California, Santa Cruz. 
\bibitem[1994]{Gou94} Goudfrooij, P. 1994, Ph.D. thesis, University of Amsterdam, The Netherlands
\bibitem[1998]{Gou98} Goudfrooij, P. 1998, in {\it Star Formation in Early--Type Galaxies}, ASP Conference Series 163, ed.s P. Carral and J. Cepa, 55
\bibitem[1998]{GT98} Goudfrooij, P., Trinchieri, G. 1998, A\&A, 330, 123
\bibitem[1985]{Ha85} Hamilton, D. 1985, ApJ, 297, 371
\bibitem[1980]{He80} Heckman, T.M. 1980, A\&A, 87, 152
\bibitem[1989]{K89} Kim, D.-W. 1989, ApJ, 346, 653
\bibitem[2000]{KZ00} Koprolin, W., Zeilinger, W.W. 2000, A\&A, 145, 71
\bibitem[1984]{Kor84} Kormendy, J. 1984, ApJ, 286, 116
\bibitem[2000]{Ku00} Kuntschner, H. 2000, MNRAS, 315, 184
\bibitem[2002]{Ku02} Kuntschner, H., Smith, R.J., Colless, M. et al. 2002, MNRAS, 337, 172
\bibitem[1987]{J87} Jarvis, B. 1987, AJ, 94, 30
\bibitem[1989]{JS89} Jedrzejewski, R., Schechter, P.L. 1989, AJ, 98, 147 
\bibitem[1989]{LV89} Lauberts, A., Valentijn E.A. 1989, {\it The Surface Photometry Catalogue of the ESO-Uppsala Galaxies}, ESO.
\bibitem[1996]{LR96} Leonardi, A.J, Rose, J.A. 1996, AJ, 111, 182
\bibitem[1998a]{L98a} Longhetti, M., Rampazzo, R., Bressan, A., Chiosi, C. 1998a, A\&A, 130, 251
\bibitem[1998b]{L98b} Longhetti, M., Rampazzo, R., Bressan, A., Chiosi, C. 1998b, A\&A, 130, 267
\bibitem[1999]{L99} Longhetti, M., Bressan, A., Chiosi, C., Rampazzo, R. 1999, A\&A, 345, 419
\bibitem[2000]{L00} Longhetti, M., Bressan, A., Chiosi, C., Rampazzo, R. 2000, A\&A, 353, 917
\bibitem[1994]{Lon94} Longo, G., Zaggia, S.R., Busarello, G. et al. 1994, A\&AS, 105, 433
\bibitem[1996]{Mac96} Macchetto, F., Pastoriza, M., Caon, N. et al.  1996, A\&AS, 120, 463
\bibitem[1983]{MC83} Malin, D.F., Carter, D. 1983, ApJ, 274, 534
\bibitem[1998]{Maoz98} Maoz, D., Koratkar, A., Shields, J.C. Ho., L.C., et al. 1998, AJ, 116, 55
\bibitem[2003]{Mehl03} Mehlert, D., Thomas, D., Saglia, R.P., Bender, R.,,Wegner, G. 2003, A\&A, 407, 423
\bibitem[1994]{Mich94} Michard, R., Marchal, J. 1994, A\&AS, 105, 481
\bibitem[1998]{Mo98} Morganti, R. Oosterloo, T., Tsvetanov, Z. 1998, AJ, 115, 915
\bibitem[1989]{Osterb89} Osterbrock, D., 1989, in {\it Astrophysics of Planetary
Nebulae and Active Galactic Nuclei}, University Science Books
\bibitem[1997]{Pi97} Pizzella, A., Amico,ÊP., Bertola,ÊF. et al. 1997, A\&A, 323, 349
\bibitem[1996]{PB96} Plana, H., Boulesteix, J. 1996, A\&A, 307, 391
\bibitem[1998]{Pl98} Plana, H., Boulesteix, J. Amram, Ph., Carignan, C., Mendes de Oliveira, C. 1998, A\&AS, 128, 75
\bibitem[1986]{Phi86} Phillips, M., Jenkins, C., Dopita, M., Sadler, E.M., Binette, L. 1986, AJ, 91, 1062
\bibitem[1996]{Phi96} Phillips, A.C., Illingworth, G.D., MacKnety, J.W. et al. 1996, AJ, 111, 1566
\bibitem[1997]{PS97} Prugniel, Ph., Simien, F. 1997, A\&AS, 122, 521 
\bibitem[1998]{PS98} Prugniel, Ph., Simien, F. 1998, A\&AS, 131, 287
\bibitem[2000]{PS00} Prugniel, Ph., Simien, F. 2000, A\&AS, 145, 263 
\bibitem[2002]{Puzia2002} Puzia, T.H., Saglia, R.P., Kissler-Patig, M., Maraston, C., Greggio, L. et al. 2002, A\&A, 395, 45
\bibitem[2000]{Qui00} Quillen, A.C., Bower, G.A., Stritzinger, M. 2000, ApJS, 128, 85
\bibitem[2001]{Ra01} Raimann, D., Storchi-Bergmann, T., Bica, E., Alloin, D.  2001, MNRAS, 324, 1087
\bibitem[1988]{R88} Rampazzo, R. 1988, A\&A, 204, 81 
\bibitem[2003]{R03} Rampazzo, R., Plana, H., Longhetti, M. et al. 2003, MNRAS, 343, 819
\bibitem[1991]{Ro91} Roberts, M.S., Hogg, D.E., Bregman, J.N., Forman, W.R., Jones, C. 1991, ApJS, 75, 751
\bibitem[1984]{Ro84} Rose J.A. 1984, AJ, 89, 1238
\bibitem[1985] {Ro85} Rose J.A. 1985, AJ, 90, 1927
\bibitem[1993]{S93} Saglia, R.P., Bertin, G., Bertola, F. et al. 1993, ApJ, 403, 567
\bibitem[1987]{RSA} Sandage, A.R., Tammann, G. 1987, {\it A Revised Shapley Ames Catalogue of Bright Galaxies}, Carnegie, Washington (RSA)
\bibitem[1999]{SFP99} Saraiva, M.F., Ferrari, F., Pastoriza, M.G. 1999, A\&A, 350, 339
\bibitem[1989]{Sch89} Schweizer, F., van Gorkom, G.H., Seitzer, P. 1989, ApJ, 338, 770
\bibitem[1996]{Sch96} Schweizer, F. 1996 in {\it Galaxies: Interactions and Induced Star Formation}, Saas-Fee Advanced Course 26. Lecture Notes 1996. Swiss Society for Astrophysics and Astronomy, XIV, Springer-Verlag Berlin/Heidelberg, p. 105
\bibitem[1983]{S83} Sharples, R.M., Carter, D., Hawarden, T.G., Longmore A.J. 1983, MNRAS, 202, 37 
\bibitem[1997a]{SP97a} Simien, F., Prugniel, Ph. 1997a, A\&AS, 126, 15
\bibitem[1997b]{SP97b} Simien, F., Prugniel, Ph. 1997b, A\&AS, 126, 519
\bibitem[2003]{Thom03} Thomas, D., Maraston, C., Bender, R. 2003, MNRAS, 339, 897
\bibitem[1998]{Tra98} Trager, S.C., Worthey, G., Faber, S.M.,  Burstein, D., Gozalez J.J. 1998, ApJS, 116, 1
\bibitem[2000]{Tra00} Trager, S.C., Faber, S.M., Worthey, G., Gozalez J.J. 2000, AJ, 119, 164
\bibitem[1997]{Tri97} Trinchieri, G., Noris, L., di Serego Alighieri, S. 1997, A\&A, 326, 565
\bibitem[1988]{Tu88} Tully, R.B. 1988, {\it Nearby Galaxy Catalogue}, Cambridge University Press 
\bibitem[1984]{U84}Ulrich-Demoulin M-H. et al. 1984,  ApJ, 285, 13
\bibitem[2000]{VD00} van Driel, W., Arnaboldi, M., Combes, F., Sparke, L.S. 2000, A\&AS, 141, 385
\bibitem[1986]{vG86} van Gorkom , J.H., Knapp, G.R., Raimond, E. et al. 1986, AJ, 91, 791
\bibitem[1987]{Va87} Varnas, S.R., Bertola, F., Galletta, G., Freeman, K.C., Carter, D. 1987, ApJ, 313, 69
\bibitem[1999]{Vaz99} Vazdekis A. 1999, ApJ, 513, 224
\bibitem[2001]{Ver01} V\`eron-Cetty, M.-P,  V\`eron, P. 2001, A\&A, 374, 92 
\bibitem[1991]{Voi91} Voit, G.M. 1991, ApJ 377, 158
\bibitem[1991]{Wh91} White III, R.,  Sarazin, C.L.  1991, ApJ, 367, 476
\bibitem[1986]{Wi86} Wilkinson, A., Sharples, R.M., Fosbury, R.A.E., Wallace, P.T. 1986, MNRAS, 218, 297
\bibitem[1992]{Wor92} Worthey, G. 1992, Ph.D. Thesis, University of California, Santa Cruz
\bibitem[1994]{Wor94} Worthey, G., Faber, S.M., Gonz\' alez, J.J., Burstein, D. 1994, ApJS, 94, 687
\bibitem[1997]{OW97} Worthey, G.,  Ottaviani, D.L. 1997, ApJS, 111, 377
\bibitem[2003] {WC03} Worthey, G., Collobert, M. 2003, ApJ, 586, 17
\bibitem[1984]{Wro84} Wrobel, J.M., Heeschen, D.S. 1984, ApJ, 287, 41
\bibitem[1996]{Z96} Zeilinger, W.W., Pizzella, A., Amico, P., Bertin, G. et al., 1996, A\&AS, 120, 257
\end{thebibliography}
\end{document}